\documentclass[aps,prl,twocolumn,superscriptaddress]{revtex4-2}
\usepackage{graphicx,braket,amsmath,amssymb} %natbib,epstopdf,siunitx,xspace,amssymb,dsfont,bm,xfrac,nicefrac, miller,
\usepackage{floatrow}
\newfloatcommand{capbtabbox}{table}[][\FBwidth]

\begin{document}
%MACROS
\newcommand{\Bi}{\textsuperscript{209}Bi}
\newcommand{\Sitwoeight}{\textsuperscript{28}Si}
\newcommand{\Sitwonine}{\textsuperscript{29}Si}
\newcommand{\Hethree}{\textsuperscript{3}He}
\newcommand{\Pp}{P_\mathrm{p}}
\newcommand{\Pth}{P_\mathrm{th}}
\newcommand{\Pso}{P_\mathrm{SO}}
\newcommand{\Zhi}{Z_\mathrm{hi}}
\newcommand{\Zlo}{Z_\mathrm{lo}}
\newcommand{\Zp}{Z_\mathrm{p}}
\newcommand{\Idc}{I_\mathrm{DC}}
\newcommand{\Ip}{I_\mathrm{p}}
\newcommand{\omegap}{\omega_\mathrm{p}}
\newcommand{\deltadc}{\delta_\mathrm{DC}}
\newcommand{\deltap}{\delta_\mathrm{p}}
\newcommand{\Deltap}{\Delta_\mathrm{p}}
\newcommand{\phip}{\phi_\mathrm{p}}
\newcommand{\alphap}{\alpha_\mathrm{p}}
\newcommand{\Pbright}{P(T|S)}
\newcommand{\Pdark}{P(T|\tilde{S})}
\newcommand{\degC}{$^\circ$C}

\title{Latched Detection of Zeptojoule Spin Echoes with a Kinetic Inductance Parametric Oscillator}

\author{Wyatt Vine}
\affiliation{School of Electrical Engineering and Telecommunications, UNSW Sydney, Sydney, NSW 2052, Australia}
\author{Anders Kringh{\o}j}
\affiliation{School of Electrical Engineering and Telecommunications, UNSW Sydney, Sydney, NSW 2052, Australia}
\author{Mykhailo Savytskyi}
\affiliation{School of Electrical Engineering and Telecommunications, UNSW Sydney, Sydney, NSW 2052, Australia}
\author{Daniel Parker}
\affiliation{School of Electrical Engineering and Telecommunications, UNSW Sydney, Sydney, NSW 2052, Australia}
\author{Thomas Schenkel}
\affiliation{Accelerator Technology and Applied Physics Division, Lawrence Berkeley National Laboratory, Berkeley, California 94720, USA}
\author{Brett C. Johnson}
\affiliation{School of Science, RMIT University, Melbourne, Victoria 3001}
\author{Jeffrey C. McCallum}
\affiliation{School  of  Physics,  University  of  Melbourne,  Parkville,  Victoria  3010,  Australia}
\author{Andrea Morello}
\affiliation{School of Electrical Engineering and Telecommunications, UNSW Sydney, Sydney, NSW 2052, Australia}
\author{Jarryd J. Pla}
\affiliation{School of Electrical Engineering and Telecommunications, UNSW Sydney, Sydney, NSW 2052, Australia}

\date{\today}

\begin{abstract}
When strongly pumped at twice their resonant frequency, non-linear resonators develop a high-amplitude intracavity field, a phenomenon known as parametric self-oscillations. The boundary over which this instability occurs can be extremely sharp and thereby presents an opportunity for realizing a detector. Here we operate such a device based on a superconducting microwave resonator whose non-linearity is engineered from kinetic inductance. The device indicates the absorption of low-power microwave wavepackets by transitioning to a self-oscillating state. Using calibrated wavepackets we measure the detection efficiency with zeptojoule energy wavepackets. We then apply it to measurements of electron spin resonance, using an ensemble of \Bi\ donors in silicon that are inductively coupled to the resonator. We achieve a latched-readout of the spin signal with an amplitude that is five hundred times greater than the underlying spin echoes.\end{abstract}

\maketitle

%\section{Teaser}
%Detection of zeptojoule energy microwave signals emitted by a spin ensemble using a superconducting parametric oscillator circuit.

\section{Introduction}
Over the past decade, quantum-limited parametric amplifiers operating at microwave frequencies have progressed from proof-of-principle to ubiquity within circuit quantum electrodynamics (cQED) experiments. Typically, these devices are operated in a linear regime, but several types of parametric amplifiers are explicitly non-linear, such as the Josephson bifurcation amplifier (JBA) \cite{Siddiqi2004} and the Josephson parametric oscillator (JPO) \cite{Lin2014,Krantz2016}. These devices essentially act as a microwave ``click''-detector, in that threshold detection is employed to discriminate the presence or absence of a signal. Central to the design of JBAs and JPOs is the use of Josephson junctions, which provide the non-linearity required for signal mixing. An alternative source of non-linearity is the kinetic inductance intrinsic to thin films of disordered superconductors \cite{HoEom2012}. In contrast to Josephson junction-based devices, superconducting microwave resonators engineered from high kinetic inductance materials retain high quality-factors when operated in tesla-strength magnetic fields \cite{Samkharadze2016,Kroll2019} and at elevated temperatures. This has recently inspired the development of magnetic field-compatible resonant parametric amplifiers that operate close to the quantum noise limit \cite{Parker2021,Xu2022,Khalifa2022,Vine2023}, which have a range of applications including axion detection \cite{Backes2021} and quantum computation with spin qubits \cite{Oakes2023}.

Another application of these devices is the measurement of electron spin resonance (ESR). JPAs have already been used to push noise in ESR experiments to the quantum limit, where vacuum fluctuations of the electromagnetic field dictate the spin detection sensitivity \cite{Bienfait2016a, Eichler2017}. Several other recent works have applied non-linear microwave amplifiers \cite{Budoyo2018} and qubit-based sensors \cite{Budoyo2020,Albertinale2021,Billaud2022,Wang2023} employing Josephson junctions to measurements of ESR in order to push detection sensitivities to record levels. Kinetic inductance parametric amplifiers (KIPAs) have also recently been used for ESR, where they have been demonstrated to have several advantages over JPAs \cite{Vine2023}. Due to their compatibility with moderate magnetic fields, KIPAs can serve as both the resonator for inductive detection of spin echo signals as well as the first-stage amplifier. This not only simplifies the measurement setup by obviating a separate quantum-limited amplifier, it also eliminates any insertion loss between the resonator and the first cryogenic amplifier. 

Here, we extend previous works with KIPAs by operating one as a ``click''-detector, rather than as a linear amplifier. By biasing the device near the threshold where its behaviour transitions from a linear amplifier to a parametric oscillator, the onset of parametric self-oscillations (PSO) serves as an indicator for the absorption of microwave wavepackets. To distinguish this operating regime, which has not been previously demonstrated, we refer to the device here as a kinetic inductance parametric oscillator (KIPO), in analogy to the JPO which operates under a similar principle. In the following we describe the concept of the detector, calibrate its sensitivity, and demonstrate its application in ESR measurements of an ensemble of bismuth ({\Bi}) donors in silicon (Si) that are directly coupled to the device.

\begin{figure*}
	\centering
	\includegraphics[width=\textwidth]{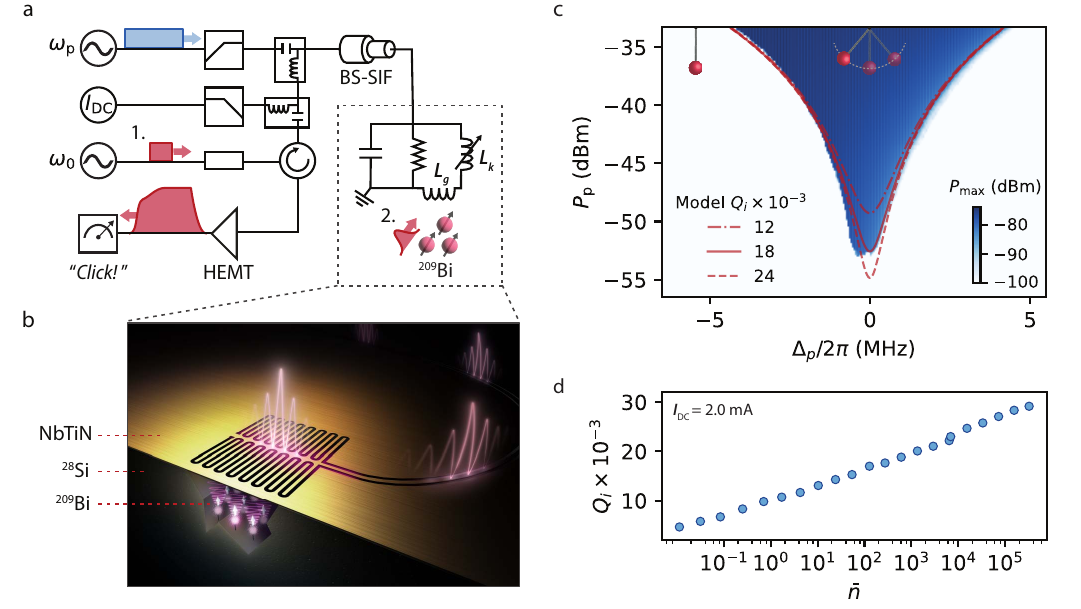}
	\caption{\textbf{Measurement schematic and device characterization.} (a) The device is represented by a parallel RLC resonator with frequency $\omega_0$ and a band-stop stepped impedance filter (BS-SIF). The resonator has a geometric inductance ($L_g$) and a kinetic inductance ($L_k$). We operate the device as a ``click''-detector by biasing it with a strong pump with frequency $\omegap$ and a DC current $\Idc$. In this work we detect two types of signals: weak classical signals with frequency near $\omega_0$ generated by a microwave source (1.), which we use to calibrate the detector's sensitivity, and spin echoes from an ensemble of {\Bi} donor spins (2.) that are resonantly coupled to the resonator via $L_g$. Detailed schematics of the measurement setups are presented in the Supplementary Material. (b) An artist's depiction of the resonator, which is formed from a $\lambda/4$ section of transmission line with a dense interdigitated capacitor. \Bi\ spins are implanted into the silicon substrate. (c) The maximum power measured with a spectrum analyzer centered at $\omega_0$ as a function of $\Deltap=\omegap-2\omega_0$ and the pump power $\Pp$. The dark blue region corresponds to the parameter space where the device self-oscillates. The powers are referred to the output of the device and $P_\mathrm{max}$ is truncated at -100~dBm to enhance clarity. The red lines correspond to the PSO threshold ($\Pth$) predicted from a model of the device using three different values of $Q_i$. The stationary and swinging pendulums are used to depict the quiet and self-oscillating states of the device, respectively. (d) $Q_i$ extracted from measurements of $S_{11}$ performed with a VNA as the signal power is varied. Measurements were taken at $T=10~$mK.}
	\label{main:fig_1}
\end{figure*}

\section{Results}

\noindent\textbf{Device Design} Our device is similar to the KIPAs described in previous works \cite{Parker2021,Vine2023}. It is patterned in a single lithographic step from a 50~nm thick film of niobium titanium nitride (NbTiN) with a kinetic inductance of $L_k=3.45~\text{pH}/\square$. The NbTiN is deposited on a Si substrate enriched in the isotope {\Sitwoeight} (750~ppm residual {\Sitwonine}) that was implanted with \Bi\ donors at a concentration of $10^{17}~\text{cm}^{-3}$ over a depth of $1.35~\mu$m. The device has a single port and  consists of a quarter-wavelength ($\lambda/4$) coplanar waveguide resonator that is shorted to ground at one end and galvanically connected to a band-stop stepped impedance filter (BS-SIF)~\cite{Bronn2015} at the other end (Figs.~\ref{main:fig_1}a,b). The resonator features a dense interdigitated capacitor with $1~\mu$m wide fingers and a $1.5~\mu$m gap to ground (see Supplementary Material), which compensates for the strong kinetic inductance and reduces the impedance of the mode to $Z_0 \approx 33~\Omega$. The BS-SIF serves to confine the resonant mode of the device while simultaneously allowing the application of a DC bias current ($\Idc$). An $\Idc$ can be used to tune the resonance frequency from $\omega_0(\Idc=0)/2\pi=7.776$~GHz to $\omega_0(\Idc=4.89~\text{mA})/2\pi=7.530$~GHz via the quadratic dependence of $L_k$ on the total current~\cite{Asfaw2017}. An $\Idc$ also enables three-wave mixing so that a pump with frequency $\omegap\approx 2\omega_0$ can be used to amplify signals with frequencies about $\omega_0$~\cite{Parker2021, Vissers2016}. The device is connected to the cold finger of a dilution refrigerator with a base temperature of 10~mK or a pumped \Hethree\ cryostat with a base temperature of 400~mK, depending on the experiment. The DC current and microwave tones are combined at base temperature using a bias tee and diplexer (Fig.~\ref{main:fig_1}a). The device is measured in reflection with the signal being routed through a cryogenic high electron mobility transistor (HEMT) amplifier at 4~K. Further details of the device design and measurement setup are provided in the Supplementary Material.

\noindent\textbf{Detector Concept} The defining characteristic of PSO is the formation of a large intracavity field at $\omega_0$ whenever the three-wave mixing strength exceeds the average rate at which resonant photons can escape the cavity. Experimentally, this manifests as the sudden generation of a large power at $\omega_0$ that is emitted from the device whenever the pump power $\Pp$ is raised beyond a sharp threshold $\Pth$. In Fig.~\ref{main:fig_1}c we report the maximum power recorded by a spectrum analyzer ($P_\mathrm{max}$) centered at $\omega_0$ as a function of $\Pp$ and a detuning of the pump frequency $\Deltap=\omegap-2\omega_0$. The dark blue region corresponds to the parameter space where the device undergoes PSO, and the boundary of this region provides a direct measurement of $\Pth$.

Using cavity input-output theory and the Hamiltonian for the KIPA~\cite{Parker2021}, we model our device as a parametrically-driven Duffing oscillator (see Supplementary Material), as has been done previously for JPOs~\cite{Wilson2010,Wustmann2013,Lin2014}. Using parameters extracted from measurements of the device we directly compare the model and our experimental measurement of the threshold power $\Pth$ (red lines in Fig.~\ref{main:fig_1}c). Crucially, the model predicts that when $\Deltap=0$, $\Pth \propto (Q_i^{-1} + Q_c^{-1})^2$, where $Q_i$ and $Q_c$ are the internal and coupling quality factors of the resonant mode, respectively. In Fig.~\ref{main:fig_1}c we compare three models of the device which are equivalent except for $Q_i$. For this particular measurement we find there is reasonable agreement between the data and model for $Q_i=18\times 10^3$. We also highlight that $\Pth$ shifts to smaller values as $Q_i$ is increased.

The quality factors $Q_i$ and $Q_c$ can be extracted from measurements of the device in reflection ($S_{11}$) using a vector network analyzer (VNA). As is common for superconducting microwave resonators, we observe that $Q_i$ is non-linear with the applied signal power, or equivalently the average number of intracavity photons $\bar{n}$ (Fig.~\ref{main:fig_1}d). This indicates that two level systems (TLSs) interact with the resonant mode~\cite{Wang2020a} and limit $Q_i$ at low signal power. We observe $Q_i$ to vary between $4.7\times 10^3$ and $29\times 10^3$ for $10^{-2} < \bar{n} < 3\times 10^5$, which in all cases is much smaller than $Q_c \approx 200\times 10^3$ (see Supplementary Material). An important consequence of this is that the resonant mode's linewidth $\omega_0(Q_i^{-1} + Q_c^{-1})$, and hence $\Pth$, are sensitive to the signal power inside the resonator.

\begin{figure}
	\centering
	\includegraphics[width=\linewidth]{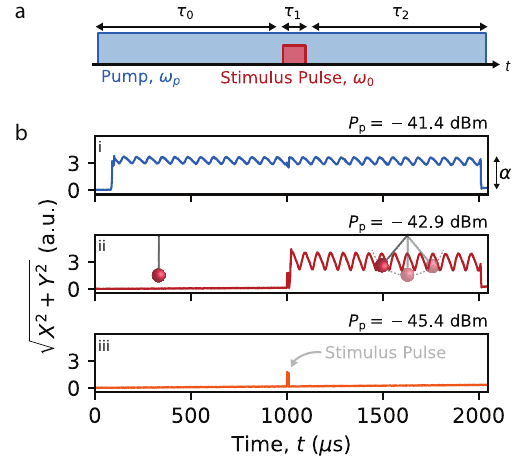}
	\caption{\textbf{Detector concept.} (a) A depiction of the two-tone pulse sequence. The device is biased with $\Idc$ and pumped at $\omegap$, before a short stimulus pulse with frequency $\omega_0$ is supplied to the device. A threshold applied to the amplitude of the demodulated signal is used to determine if the device is self-oscillating. (b) Single shots of the pulse sequence for decreasing $\Pp$ (i to iii). For certain $\Pp$, the stimulus pulse triggers the onset of PSO (ii). For these experiments, $\tau_0=\tau_2=1~$ms, $\tau_1=10~\mu$s, $\Idc = 2.55$~mA, $P_0 = -116.4$~dBm and $T=400~$mK. The time is measured from the leading edge of the pump pulse. The stimulus pulse is seen directly in (iii) at $t=1000~\mu$s.}
	\label{main:fig_2}
\end{figure}

The experiments shown in Fig.~\ref{main:fig_2} demonstrate how the dependence of $\Pth$ on $\bar{n}$ can be used to create a device that detects the absorption of low-power microwave wavepackets. First, the device is biased with a DC current $\Idc=2.55~$mA and a pump tone with frequency $\omegap=2\omega_0$. Following a delay $\tau_0=1~$ms, a stimulus pulse with duration $\tau_1=10~\mu$s, frequency $\omega_0$, and power $P_0=-116.4~$dBm is delivered to the input of the device (Fig.~\ref{main:fig_2}a). Throughout the experiment we monitor the device by performing a homodyne measurement (i.e. demodulating the signal using a local oscillator with frequency $\omega_\mathrm{LO}=\omega_0$) and recording the amplitude of the signal quadrature components $X$ and $Y$. For the lowest pump power $\Pp$ (Fig.~\ref{main:fig_2}b, iii), we observe no PSO. For the largest $\Pp$ (Fig.~\ref{main:fig_2}b, i), we observe PSO, but at a time that is uncorrelated with the stimulus pulse. But for an appropriately chosen $\Pp \approx \Pth$ (Fig.~\ref{main:fig_2}b, ii) we observe the PSO onset at a time that is correlated with the stimulus pulse (the quiet and self-oscillating states are represented schematically with the pendulum). We hypothesize that absorption of the stimulus pulse triggers the onset of PSO due to the partial saturation of the TLSs; the increased $Q_i$ associated with this results in $\Pth$ being dynamically reduced below the $\Pp$ setpoint, thereby triggering PSO. 

From Fig.~\ref{main:fig_2}b it is clear that the self-oscillating state has a large amplitude ($\alpha$), relative to the stimulus pulse. Indeed, Fig.~\ref{main:fig_1}c shows that the peak power of PSO can exceed -75~dBm. This can be understood by noting that $\alpha \propto 1/K$ for a Duffing oscillator, where $K$ is the self-Kerr strength (see Supplementary Material). The Kerr effect for these devices is known to be negligible relative to Josephson junction-based devices, due to the weak and distributed nature of the kinetic inductance non-linearity \cite{Parker2021}. By comparing the power of PSO measured with a spectrum analyzer to the average number of intracavity photons $\bar{n}$ calculated from VNA measurements of $S_{11}$ with a known power, we estimate that $\bar{n}>10^5$ during PSO. For this device, $\alpha$ is large enough to enable subsequent four wave mixing processes, which results in the generation of a frequency comb about $\omega_0$ whenever the device self-oscillates (see Supplementary Material). This results in oscillations in the amplitude of the demodulated signal when the device is self-oscillating (Fig.~\ref{main:fig_2}b).

\begin{figure*}
	\centering
	\includegraphics[width=\linewidth]{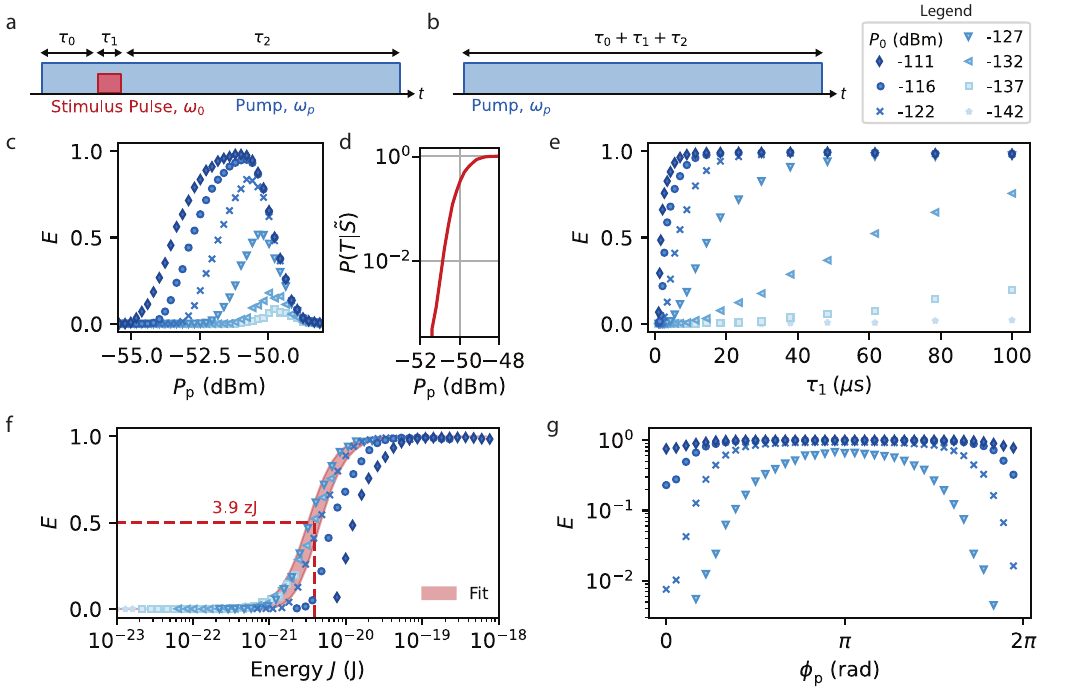}
	\caption{\textbf{Detection efficiency measured with calibrated pulses.} (a,b) The experimental pulse sequences used to measure the efficiency of the detector $E$. The sequence in panel b is a control experiment that measures dark counts. (c) $E$ measured as a function of the stimulus pulse power $P_0$ and pump power $\Pp$. (d) The probability of dark counts $\Pdark$ as a function of $\Pp$. (e) $E$ measured as a function of $P_0$ and $\tau_1$ for $\Pp=-51.2$~dBm. (f) The data in panel (e) re-plotted as a function of total wavepacket energy $J=P_0\tau_1$. We determine the lower bound of the detector's sensitivity by fitting the data with $P_0 < -116~$dBm with a sigmoid function (red shaded region). (g) $E$ measured as a function of the pump phase $\phip$ with $\tau_1=10~\mu$s. For all experiments $\Idc=2.0~$mA, $\tau_0=10~\mu$s, $\tau_2=100~\mu$s and $T=10~$mK. For each data point, at least $10^4$ shots of both the pulse sequences are measured. For all panels, the uncertainties in $E$ are too small to be seen.} 
	\label{main:fig_3}
\end{figure*}

\noindent\textbf{Detection Efficiency and Sensitivity} To calibrate the sensitivity of the detector, we measure its response using two pulse sequences. The first is depicted in Fig.~\ref{main:fig_3}a and is identical to that used in the previous section (Fig.~\ref{main:fig_2}a) but with different timings $\tau_0$, $\tau_1$ and $\tau_2$. The second sequence is similar, differing only in that the stimulus pulse is omitted (Fig.~\ref{main:fig_3}b). Using threshold detection, this allows us to measure the efficiency of the sensor, $E=\Pbright-\Pdark$, where $T$ indicates that the device self-oscillates and $S$ ($\tilde{S}$) indicates that the device receives (does not receive) a stimulus pulse. $\Pbright$ is therefore the conditional probability describing the successful detection of the stimulus and $\Pdark$ is the probability of observing a dark count. 

In Fig.~\ref{main:fig_3}c, we measure the detection efficiency $E$ as a function of the pump power $\Pp$ with a DC current $\Idc=2.0$~mA, stimulus pulse duration $\tau_1=10~\mu$s, and stimulus pulse powers $P_0$ in the range $[-137,-111]~$dBm. For these experiments we also phase modulate the pump microwave source at a rate of $15~$kHz because the gain of a linear parametric amplifier ($\Pp < \Pth$) operated in degenerate mode ($\omegap=2\omega_0$) is dependent upon the relative phase between the pump and signal ($\phip$) \cite{Parker2021}. While the modulation rate is slower than $1/\tau_1=100~$kHz, for each data point we measure $10^4$ shots of the pulse sequence to ensure the unbiased sampling of all $\phip$, thereby mitigating its effect. $E$ is found to grow monotonically with $P_0$ and is non-zero over a $7~$dB range in $\Pp$. For the largest $P_0$ measured ($P_0=-111~$dBm), $E$ reaches a maximum of 0.98, which indicates that the device functions as a near perfect detector. As $P_0$ is reduced, the $\Pp$ at which $E$ is maximized grows slightly (from $\Pp=-51.2~$dBm for $P_0=-111~$dBm to $\Pp=-49.4~$dBm for $P_0=-137~$dBm). This reflects that as $P_0$ is reduced, the device needs to be biased increasingly closer to $\Pth$ for successful detection, which results in a corresponding increase to the number of dark counts. In Fig.~\ref{main:fig_3}d we show that the probability of dark counts $\Pdark$ grows from $<0.01$ to $>0.99$ over a $3.1$~dB range in $\Pp$. This corresponds to a dark count rate that is $<0.8$~Hz for $\Pp<-51.4~$dBm and $>23~$kHz for $\Pp=-48.3~$dBm (see Supplementary Material). We note that while we chose to focus on the setpoint $\Idc=2.0~$mA in the main text, the device achieves high $E$ over a tunable range of $\sim 100~$MHz (see Supplementary Material).

Next we measure the detection efficiency $E$ as a function of the stimulus pulse duration $\tau_1$ (Fig.~\ref{main:fig_3}e). We fix the pump power $\Pp$ to the value where $E$ was maximized in the previous experiment ($-51.2$~dBm) and continue to phase modulate the pump. In this experiment the dark count probability $\Pdark$ is at maximum $1.5\times 10^{-2}$, which ensures that $E$ mainly reflects the probability of true detection $\Pbright$. $E$ reaches a maximum of $E=0.995$ and is found to grow monotonically with both $P_0$ and $\tau_1$. This suggests that $E$ is strongly correlated with the energy of the wavepacket $J=P_0\tau_1$. We confirm this by plotting $E(J)$ (Fig.~\ref{main:fig_3}f), where we see that for all $P_0$, $E(J)$ resembles an activation curve. Fitting the entire dataset with a sigmoid we infer that $E=0.5$ for $J=5.0_{4.0}^{6.3}$~zJ, where the upper and lower bounds correspond to the uncertainty in the calibration of $P_0$. We note that the $E$ achieved for a given $J$ does show some dependence on $P_0$, with the two largest $P_0$ measured having the lowest $E$. For these two measurements the cavity ring-down time is as long as $2\omega_0(Q_i^{-1} + Q_c^{-1}) \approx 1.1~\mu$s, which reduces the detection sensitivity to pulses with short $\tau_1$ at such high powers. This may explain why their activation curves (Fig.~\ref{main:fig_3}f) do not align with those taken at lower $P_0$, where $Q_i$ (and therefore the ring-down time) is reduced. Excluding the data with $P_0\geq-116~$dBm from the fit (red shaded area in Fig.~\ref{main:fig_3}f) we find the lower bound for the sensitivity to be $J=3.9_{3.1}^{4.9}$~zJ for $E=0.5$. This corresponds to wavepackets containing $756^{952}_{601}$ microwave photons, measured at the input of the device. A complementary method for determining the sensitivity of a detector is plotting the receiver operating characteristic curve (see Supplementary Material), where we find that the detector performs better than a random binary classifier for pulse energies above $0.21_{0.17}^{0.27}$~zJ ($42_{33}^{52}$ photons).

Finally, we examine the influence of the pump phase $\phip$ on the detection efficiency $E$ by turning off phase modulation on the pump microwave source and instead controlling for and stepping $\phip$ throughout the experiment (Fig.~\ref{main:fig_3}g). For each pump power $\Pp$, we found that $E$ could be both enhanced and suppressed by controlling $\phip$, relative to a control experiment where the pump microwave source was phase modulated. We also found that $E$ averaged across all $\phip$ agreed closely with the efficiency obtained with phase modulation. This confirms for the experiments in Figs.~\ref{main:fig_3}c-f that even though $1/\tau_1$ is faster than the phase modulation rate, the large number of shots taken ensures that $E$ is independent of $\phip$ when phase modulation is enabled. For $P_0=-127$~dBm (the weakest power measured in this experiment), $E$ could be suppressed to zero or made as large as 0.68, while with phase modulation $E=0.21$. The phase-dependence of $E$ highlights an essential aspect underlying the operation of this detector: the microwaves absorbed into the resonator are first parametrically amplified. The amplified signal saturates the TLSs, thereby triggerring PSO.

\begin{figure}
	\includegraphics[width=\linewidth]{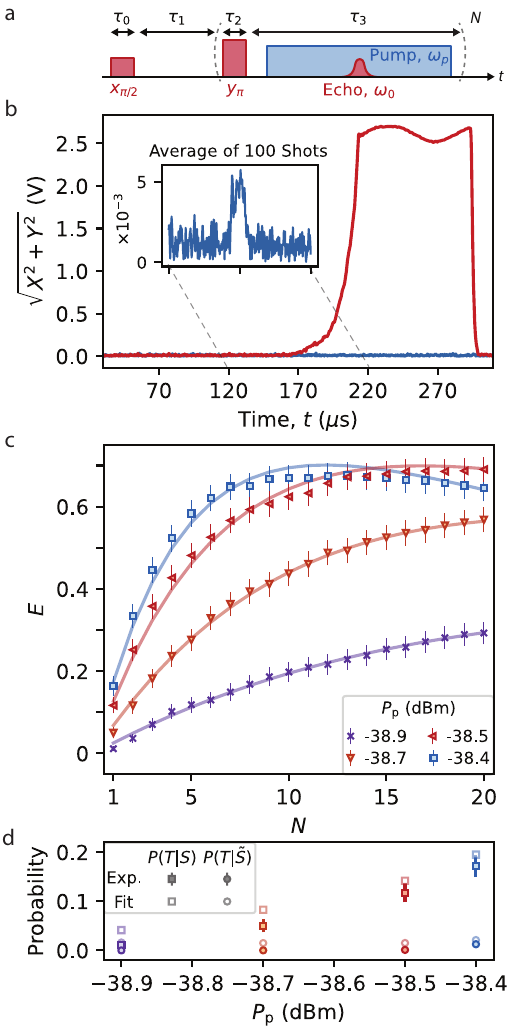}
	\caption{\textbf{Latched readout of \Bi\ spin echoes.} (a) A schematic of the Hahn echo ($N=1$) and CPMG-$N$ pulse sequences. For the pulses, $x$ and $y$ refer to the phase of the signal while $\pi/2$ and $\pi$ refer to the tipping angle. $\Pdark$ was measured using sequences where either the $x_{\pi/2}$ or $y_{\pi}$ pulses were omitted, so that no spin echo was produced. Both control sequences were performed and resulted in a similar number of dark counts. (b) Single shots of the Hahn echo pulse sequence with the pump off (blue) and with with $\Pp=-38.4~$dBm (red), measured at a field of $B_0 = 13.71$~mT. Inset: an average of 100 Hahn echoes measured with the pump off -- note the scaling of the y-axis. (c) $E$ measured as a function of $N$. For each $\Pp$, a total of $1040$ shots of the pulse sequence with $N=20$ were measured. The errorbars correspond to the 95\% confidence interval. The solid lines are fits of the data to Eq.~\ref{main:equation_efficiency_scaling}. (d) A comparison between the experimental values of $\Pbright$ and $\Pdark$ for $N=1$ and those fit to the data using Eq.~\ref{main:equation_efficiency_scaling}. All experiments were completed at $T=400~$mK and the pump was phase modulated at a rate of 15~kHz.}
	\label{main:fig_4}
\end{figure}

\noindent\textbf{Latched Readout of a Spin Ensemble} To measure ESR we apply an in-plane magnetic field of strength $B_0=13.71$~mT to bring an ESR transition of the \Bi\ donors, which were implanted into the Si substrate prior to fabrication, into resonance with the device (see Methods for a description of the spin transition measured). Resonant pulses delivered to the device are then used to control the sub-ensemble of spins with Larmor frequencies within the bandwidth of the resonator. The spins are measured using a Hahn echo pulse sequence, which is depicted in Fig.~\ref{main:fig_4}a. The first pulse in the sequence ($x_{\pi/2}$) is a $\pi/2$ pulse which causes the spins to precess about $B_0$ and dephase due to interactions with their environment. After a time delay $\tau_1$, a phase-shifted $\pi$ pulse ($y_\pi$) partially reverses the dephasing and causes the spins to refocus and emit an echo, temporarily populating the resonator with photons \cite{Morton2018}. For the Hahn echo pulse sequence this refocusing procedure is performed only once, which we designate $N = 1$ (see Fig.~\ref{main:fig_4}a). 

To detect the spins via PSO, we modify the standard Hahn echo pulse sequence by adding in a strong parametric pump following the refocusing pulse. As in the previous experiments, we set the pump power $\Pp$ close to but below the threshold power $\Pth$. PSO may then be triggered when the echo populates the resonator. Individual shots of the pulse sequence are depicted in Fig.~\ref{main:fig_4}b. When no pump tone is supplied (blue trace) the device functions simply as a resonator. The first amplifier in the measurement chain is a HEMT at 4~K. As a result, the signal-to-noise ratio (SNR) is poor and to reliably resolve the echo, the sequence must be repeated so that the signal can be averaged (inset of Fig.~\ref{main:fig_4}b). When the pump is on, the echo may trigger PSO, resulting in a detection signal with an amplitude that is a factor $\sim500$ greater than that of the echo (red trace). Moreover, while the echo itself is $\approx 10~\mu$s long, corresponding to the duration of the resonant pulses used for spin control, the PSO persists until the pump is turned off. This latched ESR readout of a spin ensemble constitutes a novel technique for detecting spin resonance.

A common technique used to enhance the SNR of pulsed ESR measurements, and thereby reduce the measurement time, is to average multiple spin echoes collected with a Carr-Purcell-Meiboom-Gill (CPMG) sequence. The CPMG pulse sequence is an extension of the Hahn-echo pulse sequence that refocuses the spins with $y_{\pi}$ pulses a total of $N$ times (Fig.~\ref{main:fig_4}a). In Fig.~\ref{main:fig_4}c we show that this technique can similarly be applied to boost the detection effieciency $E$ of our detector. For each $\Pp$ we perform 1040 shots of the pulse sequence with $N=20$ refocusing pulses. We calculate $E(N)$ by determining whether PSO were triggered for any of the $N$ repetitions within each shot of the pulse sequence. We find that by repeatedly refocusing the spin echo, $E$ can be increased by up to a factor of 25 for $\Pp=-38.9~$dBm, which in absolute terms corresponds to an improvement in $E$ by 0.28. For the three larger values of $\Pp$ measured, the absolute improvement to $E$ is even greater, with all being enhanced by more than $0.5$ for a modest number of repetitions $N$.

The experimental data in Fig.~\ref{main:fig_4}c is well-fit by the simple equation

\begin{equation}
    E(N) = [1-\Pdark]^N - [1-\Pbright]^N
    \label{main:equation_efficiency_scaling}
,\end{equation}

\noindent which is a good approximation provided the amplitude of the spin echo does not decay appreciably within the time required to complete the $N$ refocusing pulses. In Fig.~\ref{main:fig_4}d we compare the values of the true detection probability $\Pbright$ and the dark count probability $\Pdark$ fit to the data with those measured using a Hahn echo sequence ($N=1$), which shows good agreement between the two sets of values. This suggests that for the present experiment, detection of the spin echoes can be modelled as a series of statistically independent events with constant probabilities for spin echo detection and dark counts. This has important implications for optimizing the detection of spin echoes with the KIPO, namely that for finite $\Pdark$, the number of refocusing pulses $N$ required to maximize $E$ scales as $\mathcal{O}(\ln[\Pbright/\Pdark])$ (see Supplementary Material). Indeed, we see in Fig.~\ref{main:fig_4}c that for $\Pp=-38.4~$dBm where $\Pbright/\Pdark = 5.3$, $E$ could be maximized with only $N=13$ refocusing pulses. This rapid convergence on the optimal efficiency means that the CPMG detection scheme should also be effective on spin systems with shorter coherence times, as are commonly found in conventional ESR spectroscopy experiments \cite{Berliner2006}, where the number of refocusing pulses that can be applied is limited. Measuring spins with lower coherence would likely require an increase in both the excitation and detection bandwidth relative to the present device.

\section{Discussion} 

The sensitivity of the detector could be improved beyond what is demonstrated here by designing a device that is critically coupled ($Q_i=Q_c$), in which case the wavepacket would be more efficiently absorbed by the device. For the current detector, $Q_c/Q_i \approx 24$ for $\bar{n}\approx 1$, in which case the fraction of power that is absorbed is only $1-|S_{11}|^2\approx 0.15$. Reducing $Q_c$ to achieve critical coupling, e.g. by modifying the BS-SIF, could therefore increase its sensitivity by nearly an order of magnitude, while simultaneously increasing the bandwidth. We also note that the peak sensitivity was measured in an experiment utilizing phase modulation of the pump microwave source, while in Fig.~\ref{main:fig_3}g we have already demonstrated that $E$ can be increased by more than a factor of three, provided the pump and signal are phase-matched.

The sensitivity of future devices could be further enhanced by increasing the strength with which the device couples to TLSs. While this approach is counter to most cQED experiments, in the present case, it will further increase the sensitivity of $Q_i$ to the average number of intracavity photons $\bar{n}$, which is central to signal detection. This could be achieved by reducing the widths of the coplanar waveguide gap ($g$) and conductor ($w$) below the dimensions of the present device ($g=1.5~\mu$m, $w=1~\mu$m). Because TLSs reside at the dielectric interface, this will have the effect of concentrating the resonant mode volume in the material hosting the TLSs \cite{Niepce2020, Wang2020a}. In addition, dielectrics with higher concentrations of TLSs \cite{Martinis2005,Gao2008,Ungerer2023} could be intentionally deposited to improve both the detection sensitivity and bandwidth. 

The experiments presented in Fig.~\ref{main:fig_4} are a proof-of-concept that demonstrate how the detector might be used in ESR experiments. While the experiments in Fig.~\ref{main:fig_4} are taken at a single magnetic field $B_0$, we also compare the conventional ESR and KIPO methods over a range of $B_0$ fields, simulating how the KIPO might be used as a spectroscopic probe (Supplementary Material). This experiment further highlights that while detection of the spin echoes with the KIPO occurs with probability $<1$, the KIPO signal is greater than conventional ESR methods even when averaging over all shots and accounting for dark counts. This is directly related to the large power of the PSO signal, which we show in Fig.~\ref{main:fig_1}c can exceed $-75~$dBm at its peak, referred to the output of the device. This is $\sim 20~$dB greater than the equivalent room-temperature Johnson-Nyquist noise measured over a bandwidth of $100~$MHz, which means that it would be possible to measure ``clicks'' produced by the KIPO even without a cryogenic HEMT amplifier.

The experiments shown in Figs.~\ref{main:fig_4}c,d show that ESR detection using the KIPO benefits from the extension of conventional ESR techniques. Further, it was recently demonstrated how ``click''-detectors can be used to perform a full-suite of ESR techniques \cite{Albertinale2021}, which may be similarly adapted to the KIPO. Continuous-wave ESR detection might also be explored, as demonstrated using classical oscillator circuits that operate deep in the oscillator regime \cite{Anders2012}. In comparing the KIPO to works using Josephson junction-based devices, we emphasize the simplicity of our experiment. The KIPO does not require the operation of any qubits for detecting the microwave signals and can be operated at elevated temperatures and directly in a magnetic field (here 400~mK and 13.71~mT, which would preclude the use of aluminum Josephson junctions). Ultimately, this enables the KIPO to be coupled directly to the spin ensemble, which obviates a following quantum-limited amplifier or a cryogenic HEMT amplifier, as noted above. Aside from simplifying near quantum-noise-limited ESR setups, these experiments also suggest that the probability of detecting spins with the KIPO may scale exponentially with the number of measurements $N$. This might be exploited in some measurements to increase the speed of spin detection relative to conventional techniques that utilize linear quantum-noise-limited amplifiers, where the SNR grows as $\sqrt{N}$. Future work will focus on comparing both detection techniques and further exploring the advantages of latched echo readout in ESR spectroscopy.

%This has the potential to reduce the measurement time required to detect spins with a given efficiency, relative to conventional techniques where the SNR grows as $\sqrt{N}$. While beyond the scope of the present work, a direct comparison of the spin sensitivity that can be achieved with the KIPO and its scaling with $N$ to measurements utilizing a linear quantum-limited parametric amplifier will be critical in establishing whether the KIPO can be used as a practical means to improve ESR measurements.

\section{Materials and Methods}

\textbf{Device Fabrication}. The silicon substrate is enriched in the isotope \Sitwoeight\, with a residual \Sitwonine\ concentration of 750~ppm. \Bi\ ions with multiple energies were implanted to a concentration of $10^{17}$~cm\textsuperscript{-3} between 0.75~$\mu$m and 1.75~$\mu$m depth and electrically activated with a 20~min anneal at 800~$^\circ$C in a N$_2$ environment. A 50~nm film of NbTiN was then sputtered (STAR Cryoelectronics). The film was determined to have a kinetic inductance of 3.45~pH/$\square$ by matching the resonant frequencies of capacitively coupled $\lambda/4$ resonators to microwave simulations (Sonnet). The device was patterned in a single step with electron beam lithography and subsequently dry-etched with a CF\textsubscript{4}/Ar plasma. The device was then mounted and wire bonded to a printed circuit board and enclosed in a 3D copper cavity.

\textbf{Measurement Setups}. The measurements are performed in two different cryogenic systems: a dry dilution refrigerator with a base temperature of 10~mK, and a pumped \textsuperscript{3}He cryostat with a base temperature of 400~mK. In both setups DC and microwave signals are combined at the coldest stage with a bias-tee and diplexer and two cryogenic circulators are used to route the reflected microwave signals to cryogenic HEMT amplifiers. In some experiments additional amplification and filtering of the microwave and baseband signals is performed at room temperature. Time resolved measurements are performed via homodyne detection with $\omega_{\mathrm{LO}}=\omegap/2$ utilizing a homemade microwave bridge. The Supplementary Material includes detailed schematics of both setups.

\noindent\textbf{VNA Measurements}. VNA measurements were fit to cavity input-output theory \cite{Probst2015a} in order to extract $\omega_0$, $Q_i$ and $Q_c$. For the data in Fig.~\ref{main:fig_1}d, a baseline was measured and subtracted from each measurement, by setting $\Idc=0$~mA, which shifted $\omega_0$ outside the measurement window.

\noindent\textbf{Measurement of $E$}. The measurements of $E$ presented in Fig.~\ref{main:fig_3} were collected by repeating each pulse sequence 500 shots at a time, with a 33~Hz repetition rate (pump duty cycle $\approx 1/250$). In each experiment, the parameters being swept ($P_0$, $\Pp$, $\tau_1$, $\phip$, and the corresponding control experiments) were selected in pseudo-random order. For each datapoint a minimum of $10^4$ shots of both the experimental and control sequences were run. A threshold applied to the amplitude of the demodulated signal $\sqrt{X^2+Y^2}$ was used to determine if the device self-oscillated. For the experiments in Figs.~\ref{main:fig_3}a-d the pump was phase-modulated at a rate of 15~kHz to ensure $E$ would not be dependent on $\phip$. In Fig.~\ref{main:fig_3}e phase modulation was disabled. Despite a 3~GHz clock used to sync the two microwave sources, a small phase drift was still evident in $\Pbright$ (the control experiments used for calculating $\Pdark$ showed no trend, as expected). To correct for the phase drift, $\Pbright$ for each experimental repetition (500~shots for each $\phip$) was fit with a phenomenological function and aligned with the minimum $E$ on $\phip=0$. For all measurements of $E$, the uncertainties were found by adding in quadrature the 95\% confidence intervals for the binomnial distributions of $\Pbright$ and $\Pdark$.

\noindent\textbf{Measurements of \Bi}. All measurements of \Bi\ were performed at 400~mK. The device is mounted so that $B_0$ is aligned parallel to the long axis of the resonator. This orientation is chosen so that the magnetic field of the resonant mode $B_1$ is perpendicular to $B_0$ for spins located underneath the resonator. Numerically solving the \Bi\ spin Hamiltonian allows one to target specific \Bi\ ESR transitions by adjusting $B_0$ until the spins are resonant with the cavity. In this work we measure the $\bra{-,4}S_x\ket{+,5}$ transition. The small discrepancy between the $B_0$ at which we measure ESR ($13.71~$mT) and the numerically-predicted field ($13.54~$mT) might be attributed to strain caused by the different coefficients of thermal expansion of the Si substrate and NbTiN thin film \cite{Pla2018a} or a slight miscalibration of the superconducting solenoid used to generate $B_0$. For the Hahn echo and CMPG sequences used for the experiments in Fig.~\ref{main:fig_4}, we use $\tau_0=\tau_2=10~\mu$s, $\tau_1=150~\mu$s, and $\tau_3= 2\tau_1 + 4\tau_0/\pi$ \cite{Slichter1996}. The leading edge and trailing edges of the pump pulses are padded by $50~\mu$s and $30~\mu$s with respect to the $\pi_y$ pulses, to avoid their amplification. The sequences were measured with a repetition time of 7~s. For the experiment in Fig.~\ref{main:fig_4}c a total of 1040~shots of the full sequence with $N=20$ were measured for both the control and experimental sequences at each $\Pp$, while alternating between the control and experimental sequences. We performed two independent control sequences: one where the $x_{\pi/2}$ pulse was excluded and a second where the twenty $y_{\pi}$ pulses were excluded. The two values of $\Pdark$ were found to agree closely with one another, despite the different timing of the pulses relative to the onset of the pump. This indicates that the dark count rate was not influenced by residual fields of the $x_{\pi/2}$ and $y_{\pi}$ pulses in this experiment. Histograms of the detector counts for the experimental and control sequences as a function of time and repetition $N$ are provided in the Supplementary Material.

\noindent\textbf{Calibration of Powers}. All powers mentioned throughout the text are referred to the input of the device enclosure. To calibrate these powers we performed three separate cool-downs of the cryostats with two additional high-frequency lines ($L_1$ and $L_2$). In the first two cool-downs, $L_1$ or $L_2$ were connected in place of the device, and $S_{21}$ measurements of each pair of lines were taken. In the third cool-down, an $S_{21}$ measurement was taken where $L_1$ and $L_2$ were connected to one another at the base temperature plate. The combination of $S_{21}$ measurements taken over the three cool-downs allowed for a full reconstruction of the gain (loss) of each line, which agreed closely with the designed amplification (attenuation). We estimate the powers to be accurate to within $\pm1$~dB.

\subsection{Acknowledgments}
J.J.P. was supported by an Australian Research Council Discovery Early Career Research Award (DE190101397) when part of this work was performed. J.J.P. and A.M. acknowledge support from the Australian Research Council Discovery Program (DP210103769). A.M. is supported by the Australian Department of Industry, Innovation and Science (Grant No. AUS-MURI000002). W.V. acknowledges financial support from Sydney Quantum Academy, Sydney, NSW, Australia. A.K. acknowledges support from the Carlsberg Foundation. T.S. was supported by the Office of Fusion Energy Sciences, U.S. Department of Energy, under contract no. DE-AC02-05CH11231. D.P. was supported by an Australian Government Research Training Program (RTP) Scholarship. The authors acknowledge support from the NSW Node of the Australian National Fabrication Facility. We acknowledge access and support to NCRIS facilities (ANFF and the Heavy Ion Accelerator Capability) at the Australian National University. We thank Robin Cantor and STAR Cryoelectronics for sputtering the NbTiN film. The authors thank Mark Johnson, Patrice Bertet, Klaus M{\o}lmer and Tim Duty for helpful discussions.

\subsection{Author contributions}
W.V. and A.K. performed the experiments. W.V. analyzed the data. M.S. and J.J.P. designed the device and W.V. and J.J.P. fabricated it. T.S. provided the isotopically enriched silicon substrate and B.C.J. and J.C.M. performed the \Bi\ implantation. D.P. helped with the measurement electronics. A.M. and J.J.P. supervised the project. W.V. and J.J.P. wrote the manuscript with input from all authors.   

\subsection{Additional information}
J.J.P. and M.S. are inventors on a patent related to this work (AU2020347099) filed by the University of New South Wales with a priority date of 09 September 2019. The authors declare that they have no other competing interests. Correspondence and requests for materials should be addressed to J.J.P.

\clearpage
%\printbibliography

\bibliographystyle{unsrtnat}
\bibliography{main}

\begin{thebibliography}{35}
\providecommand{\natexlab}[1]{#1}
\providecommand{\url}[1]{\texttt{#1}}
\expandafter\ifx\csname urlstyle\endcsname\relax
  \providecommand{\doi}[1]{doi: #1}\else
  \providecommand{\doi}{doi: \begingroup \urlstyle{rm}\Url}\fi

\bibitem[Siddiqi et~al.(2004)Siddiqi, Vijay, Pierre, Wilson, Metcalfe, Rigetti,
  Frunzio, and Devoret]{Siddiqi2004}
I.~Siddiqi, R.~Vijay, F.~Pierre, C.~M. Wilson, M.~Metcalfe, C.~Rigetti,
  L.~Frunzio, and M.~H. Devoret.
\newblock {RF-driven Josephson bifurcation amplifier for quantum measurement}.
\newblock \emph{Physical Review Letters}, 93\penalty0 (20):\penalty0 207002,
  2004.
\newblock ISSN 00319007.
\newblock \doi{10.1103/PhysRevLett.93.207002}.

\bibitem[Lin et~al.(2014)Lin, Inomata, Koshino, Oliver, Nakamura, Tsai, and
  Yamamoto]{Lin2014}
Z.~R. Lin, K.~Inomata, K.~Koshino, W.~D. Oliver, Y.~Nakamura, J.~S. Tsai, and
  T.~Yamamoto.
\newblock {Josephson parametric phase-locked oscillator and its application to
  dispersive readout of superconducting qubits}.
\newblock \emph{Nature Communications}, 5:\penalty0 4480, 2014.
\newblock ISSN 20411723.
\newblock \doi{10.1038/ncomms5480}.

\bibitem[Krantz et~al.(2016)Krantz, Bengtsson, Simoen, Gustavsson, Shumeiko,
  Oliver, Wilson, Delsing, and Bylander]{Krantz2016}
Philip Krantz, Andreas Bengtsson, Micha{\"{e}}l Simoen, Simon Gustavsson,
  Vitaly Shumeiko, W.~D. Oliver, C.~M. Wilson, Per Delsing, and Jonas Bylander.
\newblock {Single-shot read-out of a superconducting qubit using a Josephson
  parametric oscillator}.
\newblock \emph{Nature Communications}, 7:\penalty0 11417, 2016.
\newblock ISSN 20411723.
\newblock \doi{10.1038/ncomms11417}.

\bibitem[{Ho Eom} et~al.(2012){Ho Eom}, Day, Leduc, and Zmuidzinas]{HoEom2012}
Byeong {Ho Eom}, Peter~K. Day, Henry~G. Leduc, and Jonas Zmuidzinas.
\newblock {A wideband, low-noise superconducting amplifier with high dynamic
  range}.
\newblock \emph{Nature Physics}, 8\penalty0 (8):\penalty0 623--627, 2012.
\newblock ISSN 17452481.
\newblock \doi{10.1038/nphys2356}.

\bibitem[Samkharadze et~al.(2016)Samkharadze, Bruno, Scarlino, Zheng,
  Divincenzo, Dicarlo, and Vandersypen]{Samkharadze2016}
N~Samkharadze, A~Bruno, P~Scarlino, G~Zheng, D~P Divincenzo, L~Dicarlo, and
  L~M~K Vandersypen.
\newblock {High-Kinetic-Inductance Superconducting Nanowire Resonators for
  Circuit QED in a Magnetic Field}.
\newblock \emph{Physical Review Applied}, 5\penalty0 (4):\penalty0 1--7, 2016.
\newblock ISSN 09601295.
\newblock \doi{10.1017/S0960129511000703}.
\newblock URL
  \url{https://journals.aps.org/prapplied/abstract/10.1103/PhysRevApplied.5.044004}.

\bibitem[Kroll et~al.(2019)Kroll, Borsoi, van~der Enden, Uilhoorn, de~Jong,
  Quintero-P{\'{e}}rez, van Woerkom, Bruno, Plissard, Car, Bakkers, Cassidy,
  and Kouwenhoven]{Kroll2019}
J.G. Kroll, F.~Borsoi, K.L. van~der Enden, W.~Uilhoorn, D.~de~Jong,
  M.~Quintero-P{\'{e}}rez, D.J. van Woerkom, A.~Bruno, S.R. Plissard, D.~Car,
  E.P.A.M. Bakkers, M.C. Cassidy, and L.P. Kouwenhoven.
\newblock {Magnetic-Field-Resilient Superconducting Coplanar-Waveguide
  Resonators for Hybrid Circuit Quantum Electrodynamics Experiments}.
\newblock \emph{Physical Review Applied}, 11\penalty0 (6):\penalty0 064053,
  2019.
\newblock ISSN 2331-7019.
\newblock \doi{10.1103/PhysRevApplied.11.064053}.
\newblock URL \url{https://link.aps.org/doi/10.1103/PhysRevApplied.11.064053}.

\bibitem[Parker et~al.(2022)Parker, Savytskyi, Vine, Laucht, Duty, Morello,
  Grimsmo, and Pla]{Parker2021}
Daniel~J Parker, Mykhailo Savytskyi, Wyatt Vine, Arne Laucht, Timothy Duty,
  Andrea Morello, Arne~L Grimsmo, and Jarryd~J Pla.
\newblock {Degenerate Parametric Amplification via Three-Wave Mixing Using
  Kinetic Inductance}.
\newblock \emph{Physical Review Applied}, 17:\penalty0 034064, 2022.

\bibitem[Xu et~al.(2023)Xu, Cheng, Wu, Liu, and Tang]{Xu2022}
Mingrui Xu, Risheng Cheng, Yufeng Wu, Gangqiang Liu, and Hong~X. Tang.
\newblock {Magnetic field-resilient quantum-limited parametric amplifier}.
\newblock \emph{PRX Quantum}, 4:\penalty0 010322, 2023.
\newblock URL \url{http://arxiv.org/abs/2209.13652}.

\bibitem[Khalifa and Salfi(2023)]{Khalifa2022}
M~Khalifa and J~Salfi.
\newblock {Nonlinearity and Parametric Amplification of Superconducting
  Nanowire Resonators in Magnetic Field}.
\newblock \emph{Physical Review Applied}, 19:\penalty0 034024, 2023.

\bibitem[Vine et~al.(2023)Vine, Savytskyi, Parker, Slack-Smith, Schenkel,
  McCallum, Johnson, Morello, and Pla]{Vine2023}
Wyatt Vine, Mykhailo Savytskyi, Daniel Parker, James Slack-Smith, Thomas
  Schenkel, Jeffrey~C. McCallum, Brett~C. Johnson, Andrea Morello, and
  Jarryd~J. Pla.
\newblock {In-situ amplification of spin echoes within a kinetic inductance
  parametric amplifier}.
\newblock \emph{Science Advances}, 9\penalty0 (10):\penalty0 adg1593, nov 2023.
\newblock \doi{10.1126/sciadv.adg1593}.
\newblock URL \url{http://arxiv.org/abs/2211.11333}.

\bibitem[Backes et~al.(2021)Backes, Palken, {Al Kenany}, Brubaker, Cahn,
  Droster, Hilton, Ghosh, Jackson, Lamoreaux, Leder, Lehnert, Lewis, Malnou,
  Maruyama, Rapidis, Simanovskaia, Singh, Speller, Urdinaran, Vale, van
  Assendelft, van Bibber, and Wang]{Backes2021}
Kelly~M. Backes, D.~A. Palken, S.~{Al Kenany}, B.~M. Brubaker, S.~B. Cahn,
  A.~Droster, Gene~C. Hilton, Sumita Ghosh, H.~Jackson, S.~K. Lamoreaux, A.~F.
  Leder, K.~W. Lehnert, S.~M. Lewis, M.~Malnou, R.~H. Maruyama, N.~M. Rapidis,
  M.~Simanovskaia, Sukhman Singh, D.~H. Speller, I.~Urdinaran, Leila~R. Vale,
  E.~C. van Assendelft, K.~van Bibber, and H.~Wang.
\newblock {A quantum-enhanced search for dark matter axions}.
\newblock \emph{Nature}, 590:\penalty0 238--242, 2021.
\newblock ISSN 23318422.
\newblock \doi{10.1038/s41586-021-03226-7}.
\newblock URL \url{http://dx.doi.org/10.1038/s41586-021-03226-7}.

\bibitem[Oakes et~al.(2023)Oakes, Ciriano-Tejel, Wise, Fogarty, Lundberg,
  Lain{\'{e}}, Schaal, Martins, Ibberson, Hutin, Bertrand, Stelmashenko,
  Robinson, Ibberson, Hashim, Siddiqi, Lee, Vinet, Smith, Morton, and
  Gonzalez-Zalba]{Oakes2023}
G.~A. Oakes, V.~N. Ciriano-Tejel, D.~F. Wise, M.~A. Fogarty, T.~Lundberg,
  C.~Lain{\'{e}}, S.~Schaal, F.~Martins, D.~J. Ibberson, L.~Hutin, B.~Bertrand,
  N.~Stelmashenko, J.~W.A. Robinson, L.~Ibberson, A.~Hashim, I.~Siddiqi,
  A.~Lee, M.~Vinet, C.~G. Smith, J.~J.L. Morton, and M.~F. Gonzalez-Zalba.
\newblock {Fast High-Fidelity Single-Shot Readout of Spins in Silicon Using a
  Single-Electron Box}.
\newblock \emph{Physical Review X}, 13\penalty0 (1):\penalty0 1--26, 2023.
\newblock ISSN 21603308.
\newblock \doi{10.1103/PhysRevX.13.011023}.

\bibitem[Bienfait et~al.(2015)Bienfait, Pla, Kubo, Stern, Zhou, Lo, Weis,
  Schenkel, Thewalt, Vion, Esteve, Julsgaard, M{\o}lmer, Morton, and
  Bertet]{Bienfait2016a}
A.~Bienfait, J.~J. Pla, Y.~Kubo, M.~Stern, X.~Zhou, C.~C. Lo, C.~D. Weis,
  T.~Schenkel, M.~L.W. Thewalt, D.~Vion, D.~Esteve, B.~Julsgaard, K.~M{\o}lmer,
  J.~J.L. Morton, and P.~Bertet.
\newblock {Reaching the quantum limit of sensitivity in electron spin
  resonance}.
\newblock \emph{Nature Nanotechnology}, 11\penalty0 (3):\penalty0 253--257,
  2015.
\newblock ISSN 17483395.
\newblock \doi{10.1038/nnano.2015.282}.

\bibitem[Eichler et~al.(2017)Eichler, Sigillito, Lyon, and Petta]{Eichler2017}
C.~Eichler, A.~J. Sigillito, S.~A. Lyon, and J.~R. Petta.
\newblock {Electron Spin Resonance at the Level of 104 Spins Using Low
  Impedance Superconducting Resonators}.
\newblock \emph{Physical Review Letters}, 118\penalty0 (3):\penalty0 037701,
  2017.
\newblock ISSN 10797114.
\newblock \doi{10.1103/PhysRevLett.118.037701}.

\bibitem[Budoyo et~al.(2018)Budoyo, Kakuyanagi, Toida, Matsuzaki, Munro,
  Yamaguchi, and Saito]{Budoyo2018}
Rangga~P. Budoyo, Kosuke Kakuyanagi, Hiraku Toida, Yuichiro Matsuzaki,
  William~J. Munro, Hiroshi Yamaguchi, and Shiro Saito.
\newblock {Electron paramagnetic resonance spectroscopy of Er3+: Y2SiO5 using a
  Josephson bifurcation amplifier: Observation of hyperfine and quadrupole
  structures}.
\newblock \emph{Physical Review Materials}, 2:\penalty0 011403, 2018.
\newblock ISSN 24759953.
\newblock \doi{10.1103/PhysRevMaterials.2.011403}.

\bibitem[Budoyo et~al.(2020)Budoyo, Kakuyanagi, Toida, Matsuzaki, and
  Saito]{Budoyo2020}
Rangga~P. Budoyo, Kosuke Kakuyanagi, Hiraku Toida, Yuichiro Matsuzaki, and
  Shiro Saito.
\newblock {Electron spin resonance with up to 20 spin sensitivity measured
  using a superconducting flux qubit}.
\newblock \emph{Applied Physics Letters}, 116:\penalty0 194001, 2020.
\newblock ISSN 23318422.
\newblock \doi{10.1063/1.5144722}.

\bibitem[Albertinale et~al.(2021)Albertinale, Balembois, Billaud, Ranjan,
  Flanigan, Schenkel, Est{\`{e}}ve, Vion, Bertet, and Flurin]{Albertinale2021}
Emanuele Albertinale, L{\'{e}}o Balembois, Eric Billaud, Vishal Ranjan, Daniel
  Flanigan, Thomas Schenkel, Daniel Est{\`{e}}ve, Denis Vion, Patrice Bertet,
  and Emmanuel Flurin.
\newblock {Detecting spins with a microwave photon counter}.
\newblock \emph{Nature}, 600:\penalty0 434--438, 2021.
\newblock URL \url{http://arxiv.org/abs/2102.01415}.

\bibitem[Billaud et~al.(2022)Billaud, Balembois, Dantec, Ran{\v{c}}i{\'{c}},
  Albertinale, Bertaina, Chaneli{\`{e}}re, Goldner, Est{\`{e}}ve, Vion, Bertet,
  and Flurin]{Billaud2022}
Eric Billaud, Leo Balembois, Marianne~Le Dantec, Milos Ran{\v{c}}i{\'{c}},
  Emanuele Albertinale, Sylvain Bertaina, Thierry Chaneli{\`{e}}re, Philippe
  Goldner, Daniel Est{\`{e}}ve, Denis Vion, Patrice Bertet, and Emmanuel
  Flurin.
\newblock {Microwave fluorescence detection of spin echoes}.
\newblock \emph{arXiv}, page 2208.13586, 2022.
\newblock URL \url{http://arxiv.org/abs/2208.13586}.

\bibitem[Wang et~al.(2023)Wang, Balembois, Ran{\v{c}}i{\'{c}}, Billaud, {Le
  Dantec}, Ferrier, Goldner, Bertaina, Chaneli{\`{e}}re, Esteve, Vion, Bertet,
  and Flurin]{Wang2023}
Z.~Wang, L.~Balembois, M.~Ran{\v{c}}i{\'{c}}, E.~Billaud, M~{Le Dantec},
  A.~Ferrier, P.~Goldner, S.~Bertaina, T.~Chaneli{\`{e}}re, D.~Esteve, D.~Vion,
  P.~Bertet, and E.~Flurin.
\newblock {Single-electron spin resonance detection by microwave photon
  counting}.
\newblock \emph{Nature}, 619:\penalty0 276--281, 2023.

\bibitem[Bronn et~al.(2015)Bronn, Liu, Hertzberg, C{\'{o}}rcoles, Houck,
  Gambetta, and Chow]{Bronn2015}
Nicholas~T. Bronn, Yanbing Liu, Jared~B. Hertzberg, Antonio~D. C{\'{o}}rcoles,
  Andrew~A. Houck, Jay~M. Gambetta, and Jerry~M. Chow.
\newblock {Broadband filters for abatement of spontaneous emission in circuit
  quantum electrodynamics}.
\newblock \emph{Applied Physics Letters}, 107:\penalty0 172601, 2015.
\newblock ISSN 00036951.
\newblock \doi{10.1063/1.4934867}.

\bibitem[Asfaw et~al.(2017)Asfaw, Sigillito, Tyryshkin, Schenkel, and
  Lyon]{Asfaw2017}
A.~T. Asfaw, A.~J. Sigillito, A.~M. Tyryshkin, T.~Schenkel, and S.~A. Lyon.
\newblock {Multi-frequency spin manipulation using rapidly tunable
  superconducting coplanar waveguide microresonators}.
\newblock \emph{Applied Physics Letters}, 111\penalty0 (3), 2017.
\newblock ISSN 00036951.
\newblock \doi{10.1063/1.4993930}.

\bibitem[Vissers et~al.(2016)Vissers, Erickson, Ku, Vale, Wu, Hilton, and
  Pappas]{Vissers2016}
M.~R. Vissers, R.~P. Erickson, H.~S. Ku, Leila Vale, Xian Wu, G.~C. Hilton, and
  D.~P. Pappas.
\newblock {Low-noise kinetic inductance traveling-wave amplifier using
  three-wave mixing}.
\newblock \emph{Applied Physics Letters}, 108\penalty0 (1), 2016.
\newblock ISSN 00036951.
\newblock \doi{10.1063/1.4937922}.

\bibitem[Wilson et~al.(2010)Wilson, Duty, Sandberg, Persson, Shumeiko, and
  Delsing]{Wilson2010}
C.~M. Wilson, T.~Duty, M.~Sandberg, F.~Persson, V.~Shumeiko, and P.~Delsing.
\newblock {Photon generation in an electromagnetic cavity with a time-dependent
  boundary}.
\newblock \emph{Physical Review Letters}, 105\penalty0 (23):\penalty0 1--4,
  2010.
\newblock ISSN 00319007.
\newblock \doi{10.1103/PhysRevLett.105.233907}.

\bibitem[Wustmann and Shumeiko(2013)]{Wustmann2013}
Waltraut Wustmann and Vitaly Shumeiko.
\newblock {Parametric resonance in tunable superconducting cavities}.
\newblock \emph{Physical Review B - Condensed Matter and Materials Physics},
  87\penalty0 (18):\penalty0 1--23, 2013.
\newblock ISSN 10980121.
\newblock \doi{10.1103/PhysRevB.87.184501}.

\bibitem[Wang et~al.(2020)Wang, McRae, Gao, Vissers, Brecht, Dunsworth, Pappas,
  and Mutus]{Wang2020a}
H.~Wang, C.~R.H. McRae, J.~Gao, M.~R. Vissers, T.~Brecht, A.~Dunsworth, D.~P.
  Pappas, and J.~Mutus.
\newblock {Materials loss measurements using superconducting microwave
  resonators}.
\newblock \emph{Review of Scientific Instruments}, 91\penalty0 (9), 2020.
\newblock ISSN 10897623.
\newblock \doi{10.1063/5.0017378}.

\bibitem[Morton and Bertet(2018)]{Morton2018}
John~J.L. Morton and Patrice Bertet.
\newblock {Storing quantum information in spins and high-sensitivity ESR}.
\newblock \emph{Journal of Magnetic Resonance}, 287:\penalty0 128--139, 2018.
\newblock ISSN 10960856.
\newblock \doi{10.1016/j.jmr.2017.11.015}.
\newblock URL \url{https://doi.org/10.1016/j.jmr.2017.11.015}.

\bibitem[Berliner et~al.(2006)Berliner, Eaton, and Eaton]{Berliner2006}
Lawrence~J Berliner, Sandra~S Eaton, and Gareth~R Eaton.
\newblock \emph{Distance measurements in biological systems by EPR}, volume~19.
\newblock Springer Science \& Business Media, 2006.

\bibitem[Niepce et~al.(2020)Niepce, Burnett, Latorre, and Bylander]{Niepce2020}
David Niepce, Jonathan~J Burnett, Mart{\'\i}~Gutierrez Latorre, and Jonas
  Bylander.
\newblock Geometric scaling of two-level-system loss in superconducting
  resonators.
\newblock \emph{Superconductor Science and Technology}, 33\penalty0
  (2):\penalty0 025013, 2020.

\bibitem[Martinis et~al.(2005)Martinis, Cooper, McDermott, Steffen, Ansmann,
  Osborn, Cicak, Oh, Pappas, Simmonds, and Yu]{Martinis2005}
John~M. Martinis, K.~B. Cooper, R.~McDermott, Matthias Steffen, Markus Ansmann,
  K.~D. Osborn, K.~Cicak, Seongshik Oh, D.~P. Pappas, R.~W. Simmonds, and
  Clare~C. Yu.
\newblock {Decoherence in Josephson qubits from dielectric Loss}.
\newblock \emph{Physical Review Letters}, 95\penalty0 (21):\penalty0 1--4,
  2005.
\newblock ISSN 00319007.
\newblock \doi{10.1103/PhysRevLett.95.210503}.

\bibitem[Gao et~al.(2008)Gao, Daal, Vayonakis, Kumar, Zmuidzinas, Sadoulet,
  Mazin, Day, and Leduc]{Gao2008}
Jiansong Gao, Miguel Daal, Anastasios Vayonakis, Shwetank Kumar, Jonas
  Zmuidzinas, Bernard Sadoulet, Benjamin~A Mazin, Peter~K Day, and Henry~G
  Leduc.
\newblock Experimental evidence for a surface distribution of two-level systems
  in superconducting lithographed microwave resonators.
\newblock \emph{Applied Physics Letters}, 92\penalty0 (15), 2008.

\bibitem[Ungerer et~al.(2023)Ungerer, Sarmah, Kononov, Ridderbos, Haller,
  Cheung, and Sch{\"o}nenberger]{Ungerer2023}
Jann~H Ungerer, Deepankar Sarmah, Artem Kononov, Joost Ridderbos, Roy Haller,
  Luk~Yi Cheung, and Christian Sch{\"o}nenberger.
\newblock Performance of high impedance resonators in dirty dielectric
  environments.
\newblock \emph{arXiv preprint arXiv:2302.06303}, 2023.

\bibitem[Anders et~al.(2012)Anders, Angerhofer, and Boero]{Anders2012}
Jens Anders, Alexander Angerhofer, and Giovanni Boero.
\newblock K-band single-chip electron spin resonance detector.
\newblock \emph{Journal of Magnetic Resonance}, 217:\penalty0 19--26, 2012.

\bibitem[Probst et~al.(2015)Probst, Song, Bushev, Ustinov, and
  Weides]{Probst2015a}
S~Probst, F~B Song, P~A Bushev, A~V Ustinov, and M~Weides.
\newblock {Efficient and robust analysis of complex scattering data under noise
  in microwave resonators}.
\newblock \emph{Review of Scientific Instruments}, 86:\penalty0 024706, 2015.
\newblock \doi{10.1063/1.4907935}.
\newblock URL \url{http://dx.doi.org/10.1063/1.4907935}.

\bibitem[Pla et~al.(2018)Pla, Bienfait, Pica, Mansir, Mohiyaddin, Zeng, Niquet,
  Morello, Schenkel, Morton, and Bertet]{Pla2018a}
J.~J. Pla, A.~Bienfait, G.~Pica, J.~Mansir, F.~A. Mohiyaddin, Z.~Zeng, Y.~M.
  Niquet, A.~Morello, T.~Schenkel, J.~J.L. Morton, and P.~Bertet.
\newblock {Strain-Induced Spin-Resonance Shifts in Silicon Devices}.
\newblock \emph{Physical Review Applied}, 9\penalty0 (4):\penalty0 44014, 2018.
\newblock ISSN 23317019.
\newblock \doi{10.1103/PhysRevApplied.9.044014}.
\newblock URL \url{https://doi.org/10.1103/PhysRevApplied.9.044014}.

\bibitem[Slichter(1996)]{Slichter1996}
C~P Slichter.
\newblock \emph{{Princples of Magnetic Resonance}}.
\newblock Springer, Berlin, 3rd ed. edition, 1996.

\end{thebibliography}


\begin{thebibliography}{15}
\providecommand{\natexlab}[1]{#1}
\providecommand{\url}[1]{\texttt{#1}}
\expandafter\ifx\csname urlstyle\endcsname\relax
  \providecommand{\doi}[1]{doi: #1}\else
  \providecommand{\doi}{doi: \begingroup \urlstyle{rm}\Url}\fi

\bibitem[Parker et~al.(2022)Parker, Savytskyi, Vine, Laucht, Duty, Morello,
  Grimsmo, and Pla]{Parker2021}
Daniel~J Parker, Mykhailo Savytskyi, Wyatt Vine, Arne Laucht, Timothy Duty,
  Andrea Morello, Arne~L Grimsmo, and Jarryd~J Pla.
\newblock {Degenerate Parametric Amplification via Three-Wave Mixing Using
  Kinetic Inductance}.
\newblock \emph{Physical Review Applied}, 17:\penalty0 034064, 2022.

\bibitem[Vine et~al.(2023)Vine, Savytskyi, Parker, Slack-Smith, Schenkel,
  McCallum, Johnson, Morello, and Pla]{Vine2023}
Wyatt Vine, Mykhailo Savytskyi, Daniel Parker, James Slack-Smith, Thomas
  Schenkel, Jeffrey~C. McCallum, Brett~C. Johnson, Andrea Morello, and
  Jarryd~J. Pla.
\newblock {In-situ amplification of spin echoes within a kinetic inductance
  parametric amplifier}.
\newblock \emph{Science Advances}, 9\penalty0 (10):\penalty0 adg1593, nov 2023.
\newblock \doi{10.1126/sciadv.adg1593}.
\newblock URL \url{http://arxiv.org/abs/2211.11333}.

\bibitem[Vissers et~al.(2015)Vissers, Hubmayr, Sandberg, Chaudhuri,
  Bockstiegel, and Gao]{Vissers2015}
M~R Vissers, J~Hubmayr, M~Sandberg, S~Chaudhuri, C~Bockstiegel, and J~Gao.
\newblock {Frequency-tunable superconducting resonators via nonlinear kinetic
  inductance}.
\newblock \emph{Applied Physics Letters}, 107:\penalty0 062601, 2015.
\newblock \doi{10.1063/1.4927444}.

\bibitem[Bruno et~al.(2015)Bruno, {De Lange}, Asaad, {Van Der Enden}, Langford,
  and Dicarlo]{Bruno2015}
A.~Bruno, G.~{De Lange}, S.~Asaad, K.~L. {Van Der Enden}, N.~K. Langford, and
  L.~Dicarlo.
\newblock {Reducing intrinsic loss in superconducting resonators by surface
  treatment and deep etching of silicon substrates}.
\newblock \emph{Applied Physics Letters}, 106:\penalty0 182601, 2015.
\newblock ISSN 00036951.
\newblock \doi{10.1063/1.4919761}.
\newblock URL \url{http://dx.doi.org/10.1063/1.4919761}.

\bibitem[Wang et~al.(2020)Wang, McRae, Gao, Vissers, Brecht, Dunsworth, Pappas,
  and Mutus]{Wang2020a}
H.~Wang, C.~R.H. McRae, J.~Gao, M.~R. Vissers, T.~Brecht, A.~Dunsworth, D.~P.
  Pappas, and J.~Mutus.
\newblock {Materials loss measurements using superconducting microwave
  resonators}.
\newblock \emph{Review of Scientific Instruments}, 91\penalty0 (9), 2020.
\newblock ISSN 10897623.
\newblock \doi{10.1063/5.0017378}.

\bibitem[Niepce et~al.(2019)Niepce, Burnett, and Bylander]{Niepce2019a}
David Niepce, Jonathan Burnett, and Jonas Bylander.
\newblock {High Kinetic Inductance Nb N Nanowire Superinductors}.
\newblock \emph{Physical Review Applied}, 11:\penalty0 044014, 2019.
\newblock ISSN 23317019.
\newblock \doi{10.1103/PhysRevApplied.11.044014}.

\bibitem[Wilson et~al.(2010)Wilson, Duty, Sandberg, Persson, Shumeiko, and
  Delsing]{Wilson2010}
C.~M. Wilson, T.~Duty, M.~Sandberg, F.~Persson, V.~Shumeiko, and P.~Delsing.
\newblock {Photon generation in an electromagnetic cavity with a time-dependent
  boundary}.
\newblock \emph{Physical Review Letters}, 105\penalty0 (23):\penalty0 1--4,
  2010.
\newblock ISSN 00319007.
\newblock \doi{10.1103/PhysRevLett.105.233907}.

\bibitem[Wustmann and Shumeiko(2013)]{Wustmann2013}
Waltraut Wustmann and Vitaly Shumeiko.
\newblock {Parametric resonance in tunable superconducting cavities}.
\newblock \emph{Physical Review B - Condensed Matter and Materials Physics},
  87\penalty0 (18):\penalty0 1--23, 2013.
\newblock ISSN 10980121.
\newblock \doi{10.1103/PhysRevB.87.184501}.

\bibitem[Lin et~al.(2014)Lin, Inomata, Koshino, Oliver, Nakamura, Tsai, and
  Yamamoto]{Lin2014}
Z.~R. Lin, K.~Inomata, K.~Koshino, W.~D. Oliver, Y.~Nakamura, J.~S. Tsai, and
  T.~Yamamoto.
\newblock {Josephson parametric phase-locked oscillator and its application to
  dispersive readout of superconducting qubits}.
\newblock \emph{Nature Communications}, 5:\penalty0 4480, 2014.
\newblock ISSN 20411723.
\newblock \doi{10.1038/ncomms5480}.

\bibitem[Erickson et~al.(2014)Erickson, Vissers, Sandberg, Jefferts, and
  Pappas]{Erickson2014}
R.~P. Erickson, M.~R. Vissers, M.~Sandberg, S.~R. Jefferts, and D.~P. Pappas.
\newblock {Frequency comb generation in superconducting resonators}.
\newblock \emph{Physical Review Letters}, 113\penalty0 (18):\penalty0 1--5,
  2014.
\newblock ISSN 10797114.
\newblock \doi{10.1103/PhysRevLett.113.187002}.

\bibitem[Cassidy et~al.(2017)Cassidy, Bruno, Rubbert, Irfan, Kammhuber,
  Schouten, Akhmerov, and Kouwenhoven]{Cassidy2017a}
M.~C. Cassidy, A.~Bruno, S.~Rubbert, M.~Irfan, J.~Kammhuber, R.~N. Schouten,
  A.~R. Akhmerov, and L.~P. Kouwenhoven.
\newblock {Demonstration of an ac Josephson junction laser}.
\newblock \emph{Science}, 355\penalty0 (6328):\penalty0 939--942, 2017.
\newblock ISSN 10959203.
\newblock \doi{10.1126/science.aah6640}.

\bibitem[Khan and T{\"{u}}reci(2018)]{Khan2018}
Saeed Khan and Hakan~E. T{\"{u}}reci.
\newblock {Frequency Combs in a Lumped-Element Josephson-Junction Circuit}.
\newblock \emph{Physical Review Letters}, 120\penalty0 (15):\penalty0 153601,
  2018.
\newblock ISSN 10797114.
\newblock \doi{10.1103/PhysRevLett.120.153601}.
\newblock URL \url{https://doi.org/10.1103/PhysRevLett.120.153601}.

\bibitem[Lu et~al.(2021)Lu, Chien, Cao, Lanes, Zhou, Hatridge, Khan, and
  T{\"{u}}reci]{Lu2021}
Pinlei Lu, Tzu~Chiao Chien, Xi~Cao, Olivia Lanes, Chao Zhou, Michael~J.
  Hatridge, Saeed Khan, and Hakan~E. T{\"{u}}reci.
\newblock {Nearly Quantum-Limited Josephson-Junction Frequency-Comb
  Synthesizer}.
\newblock \emph{Physical Review Applied}, 15\penalty0 (4):\penalty0 1, 2021.
\newblock ISSN 23317019.
\newblock \doi{10.1103/PhysRevApplied.15.044031}.
\newblock URL \url{https://doi.org/10.1103/PhysRevApplied.15.044031}.

\bibitem[Wang et~al.(2021)Wang, Chen, and Li]{Wang2021a}
Shuai~Peng Wang, Zhen Chen, and Tiefu Li.
\newblock {Controllable microwave frequency comb generation in a tunable
  superconducting coplanar-waveguide resonator}.
\newblock \emph{Chinese Physics B}, 30\penalty0 (4), 2021.
\newblock ISSN 20583834.
\newblock \doi{10.1088/1674-1056/abc2bb}.

\bibitem[Qiu et~al.(2022)Qiu, Grimsmo, Peng, Kannan, Lienhard, Sung, Krantz,
  Bolkhovsky, Calusine, Kim, Melville, Niedzielski, Yoder, Schwartz, Orlando,
  Siddiqi, Gustavsson, O'Brien, and Oliver]{Qiu2022}
Jack~Y. Qiu, Arne Grimsmo, Kaidong Peng, Bharath Kannan, Benjamin Lienhard,
  Youngkyu Sung, Philip Krantz, Vladimir Bolkhovsky, Greg Calusine, David Kim,
  Alex Melville, Bethany~M. Niedzielski, Jonilyn Yoder, Mollie~E. Schwartz,
  Terry~P. Orlando, Irfan Siddiqi, Simon Gustavsson, Kevin~P. O'Brien, and
  William~D. Oliver.
\newblock {Broadband Squeezed Microwaves and Amplification with a Josephson
  Traveling-Wave Parametric Amplifier}.
\newblock \emph{arXiv}, page 2201.11261, 2022.
\newblock URL \url{http://arxiv.org/abs/2201.11261}.

\end{thebibliography}

%references in supplementary

\end{document}

\title{Supplementary Material: Latched Detection of Zeptojoule Spin Echoes with a Kinetic Inductance Parametric Oscillator}

\author{Wyatt Vine}
\affiliation{School of Electrical Engineering and Telecommunications, UNSW Sydney, Sydney, NSW 2052, Australia}
\author{Anders Kringh{\o}j}
\affiliation{School of Electrical Engineering and Telecommunications, UNSW Sydney, Sydney, NSW 2052, Australia}
\author{Mykhailo Savytskyi}
\affiliation{School of Electrical Engineering and Telecommunications, UNSW Sydney, Sydney, NSW 2052, Australia}
\author{Daniel Parker}
\affiliation{School of Electrical Engineering and Telecommunications, UNSW Sydney, Sydney, NSW 2052, Australia}
\author{Thomas Schenkel}
\affiliation{Accelerator Technology and Applied Physics Division, Lawrence Berkeley National Laboratory, Berkeley, California 94720, USA}
\author{Brett C. Johnson}
\affiliation{School of Science, RMIT University, Melbourne, Victoria 3001}
\author{Jeffrey C. McCallum}
\affiliation{School  of  Physics,  University  of  Melbourne,  Parkville,  Victoria  3010,  Australia}
\author{Andrea Morello}
\affiliation{School of Electrical Engineering and Telecommunications, UNSW Sydney, Sydney, NSW 2052, Australia}
\author{Jarryd J. Pla}
\affiliation{School of Electrical Engineering and Telecommunications, UNSW Sydney, Sydney, NSW 2052, Australia}

\date{\today}
\maketitle
\tableofcontents
\pagebreak

\newcommand{\beginsupplement}{%
        \setcounter{table}{0}
        \renewcommand{\thetable}{S\arabic{table}}%
        \setcounter{figure}{0}
        \renewcommand{\thefigure}{S\arabic{figure}}%
     }
\beginsupplement

\section{Measurement Setups}
\subsection{Measurements at 10~mK}

\begin{figure}
	\centering
	\includegraphics[width=0.6\linewidth]{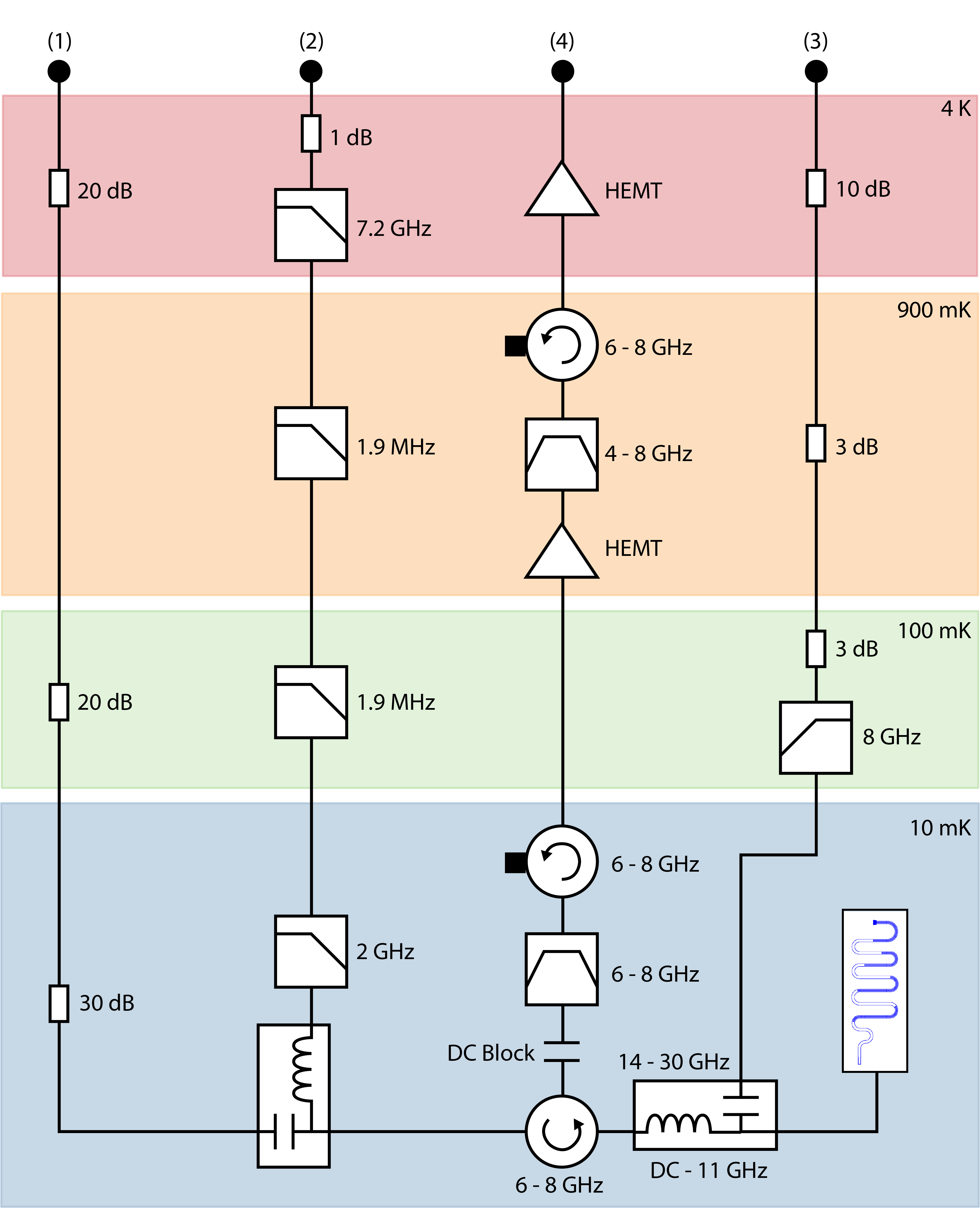}
	\caption{Schematic for measurements performed at 10 mK.} 
	\label{venus_schematic}
\end{figure}

Measurements at 10~mK were performed using a Bluefors LD-250 dilution refrigerator. The copper sample enclosure was mounted directly to the mixing chamber (MXC) plate. We utilized 4 lines in the experiment, labeled (1)-(4) in Fig.~\ref{venus_schematic}. (1) The signal line was used to supply tones near the resonance frequency of the device ($\omega_0/2\pi \approx7.7$~GHz). (2) The DC line was used to apply a DC-current to the device. (3) The pump line was used to send tones near twice the resonance frequency of the device ($\omegap/2\pi \approx15.4$~GHz). (4) The output line was used to amplify the reflected microwave signals. The signals on lines (1), (2) and (3) were combined with a bias-tee (Pasternack PE1615) and a diplexer (Marki DPX114) and fed into the device via a short semi-rigid cable connected to the common port of the diplexer.

To ensure that the noise reaching the device at $\omega_0$ was limited by the equilibrium fluctuations at the lowest temperature stage (10 mK), 70~dB of fixed broadband attenuation was used on line (1) and a series of reflective low-pass filters were used on line (2). To facilitate the application of a strong pump, line (3) had only 16~dB of fixed attenuation, but high-pass filters at the lowest temperature stages ensured $>60~$dB of attenuation for frequencies about $\omega_0$.

The reflected microwave signals were routed through two cryogenic HEMT amplifiers situated at 1~K and 4~K. To protect the amplifiers from damage and radiation from reaching the sample, a DC-block, two band-pass filters, and two isolators were used.

\subsection{Measurements at 400~mK}

Measurements at 400~mK were performed using a \textsuperscript{3}He refrigerator. The copper sample enclosure was mounted to the base-plate of the insert and enclosed in a vacuum can that was dipped directly into liquid \textsuperscript{4}He. A closed-cycle, single-shot \textsuperscript{3}He system maintained the sample at $\sim 400~$mK throughout the duration of the measurements. The attenuation and filtering of the four lines was configured similarly to that of the 10~mK measurements.

\begin{figure}
	\centering
	\includegraphics[width=0.6\linewidth]{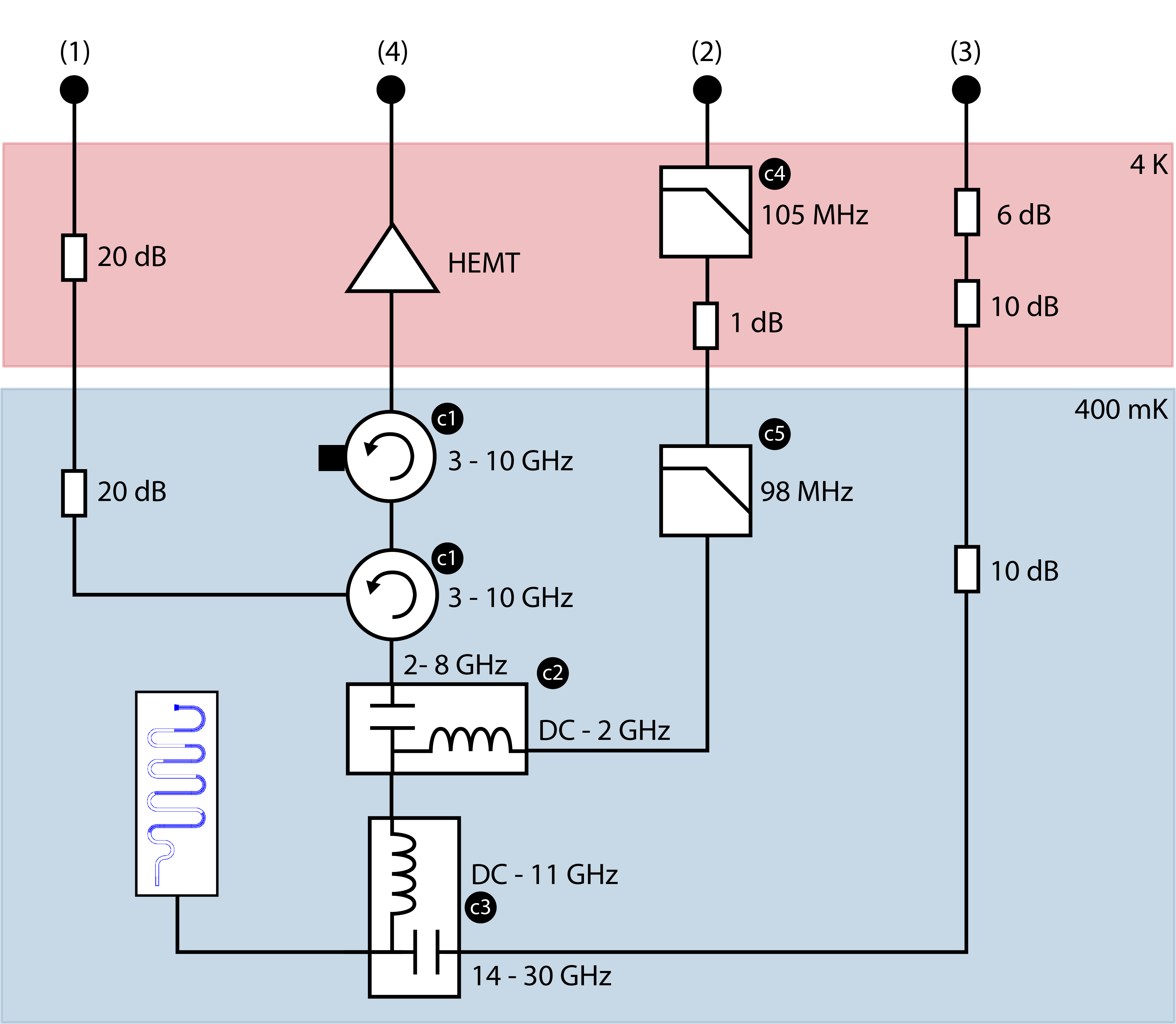}
	\caption{Schematic for measurements performed at 400 mK. The components are (c1) Raditek RADC-4-10-Cryo circulator, (c2) Marki DPXN4, (c3) Marki DPX1114, (c4) Mini-Circuits VLF-105+, (c5) Mini-Circuits SLP-100+.} 
	\label{LG07_lemon_schematic}
\end{figure}

\subsection{Microwave Bridge}

\begin{figure}
	\centering
	\includegraphics[width=0.6\linewidth]{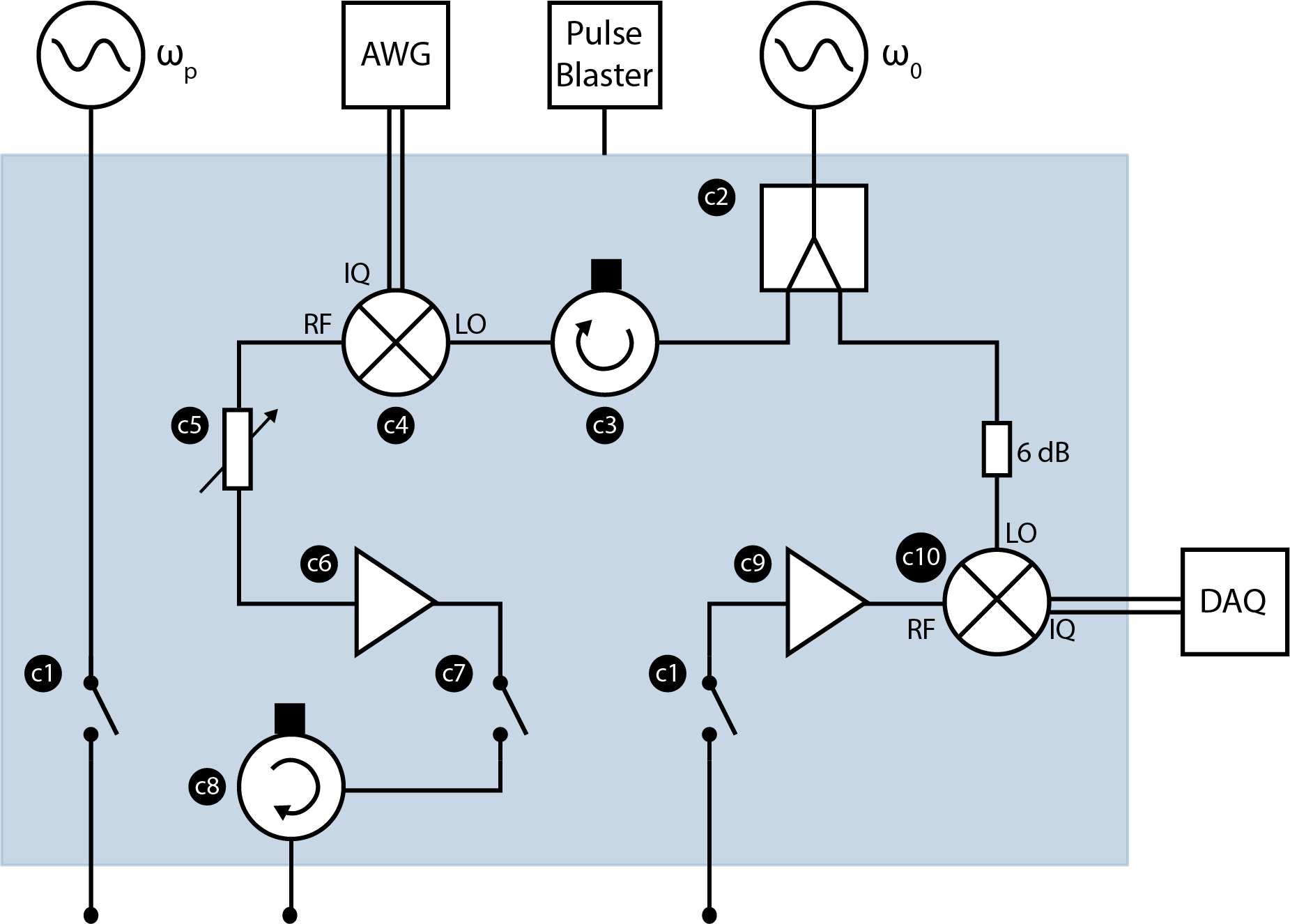}
	\caption{Schematic for the home-built spectrometer used for pulsed ESR measurements. c1: single-pole double-throw microwave switch, c2: Marki PD-0R510, c3: Ditom D3C4080, c4: Marki IQ-4509LXP, c5: Mini-Circuits RUDAT-13G-60, c6: Mini-Circuits ZVE 3W-183+, c7: double-pole quad-throw microwave switch, c9: Mini-Circuits ZX60-05113LN+, c10: Polyphase AD60100B, $\omega_0$ microwave source: Keysight PNA-L N5231B, $\omegap$ microwave source: Anritsu MG3692b, AWG: Keysight M3202A, DAQ: Keysight M3102A.} 
	\label{fig_SI:kubrick}
\end{figure}

Time-resolved experiments were performed using a homemade microwave spectrometer which is depicted in Fig.~\ref{fig_SI:kubrick}. The system has two main paths: one for generating phase-coherent pulses and one for performing homodyne demodulation of the signals coming from the device. Pulses are generated by IQ mixing a local oscillator (LO) with baseband signals from an arbitrary waveform generator (AWG). To extend the dynamic range of the system, a programmable attenuator is used to vary the output power of the system. To suppress leakage of signals from the box when no pulses are being sent, a fast microwave switch is placed before the output of the system and is actuated throughout the pulse sequences. Similar switches are placed at the input of the box and provide $>40$~dB of attenuation when open to ensure that the high power pulses used for performing ESR do not reach the demodulator with a power exceeding its damage threshold. Power entering the box is further amplified before being homodyne demodulated. The resulting quadrature signals, $X(t)$ and $Y(t)$, are digitized by a data acquisition system (DAQ). For experiments where the pump is added to a pulse sequence, it is routed through the box and gated by a fast microwave switch. Triggering of the AWG, DAQ, and microwave switches is achieved with TTL logic supplied by a pulse generator (Spin-Core PulseBlaster ESR Pro).

\section{Device Design}

\begin{figure}
	\centering
	\includegraphics[width=0.9\linewidth]{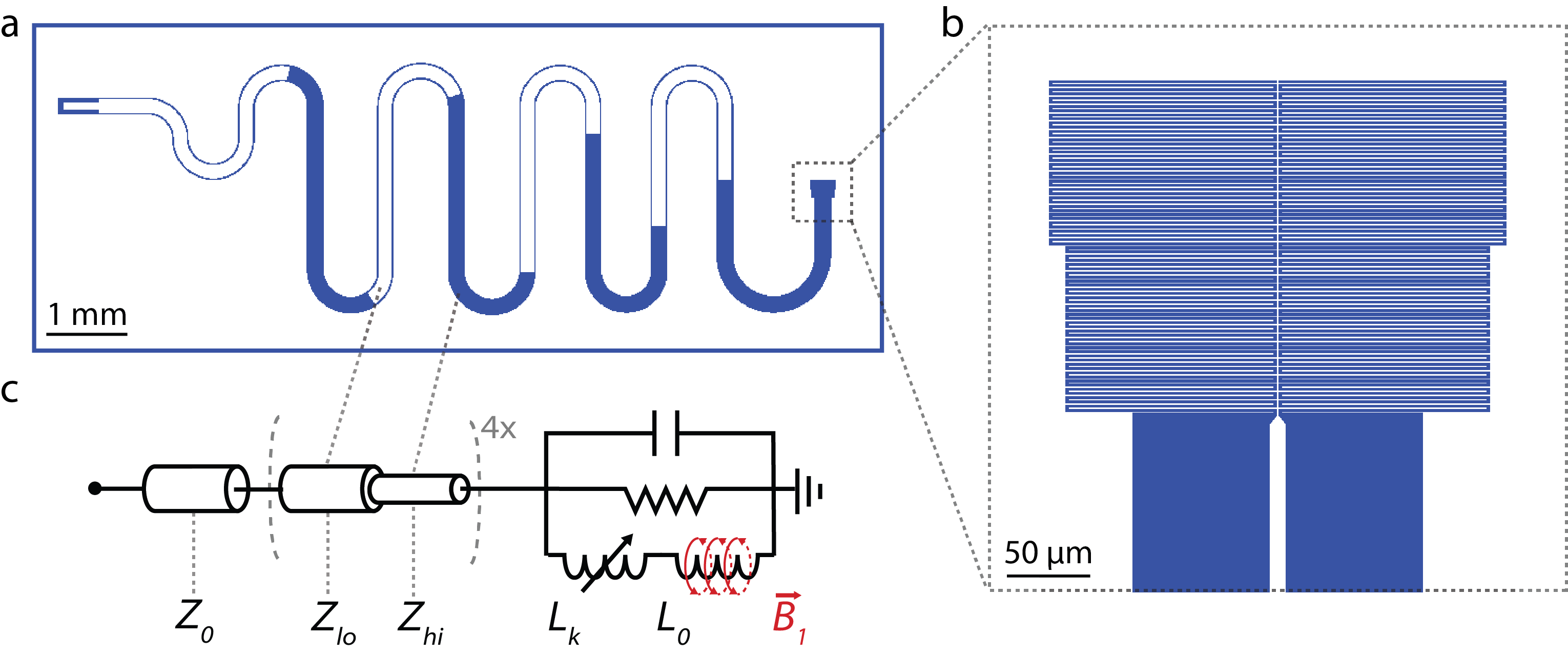}
	\caption{Device schematic. (a) A top-down view of the full device. The white regions correspond to NbTiN and the blue regions where it has been etched to reveal the silicon underneath. The large meandering CPW corresponds to the BS-SIF (note that at this scale the centre conductor of the $\Zhi$ section is too small to be seen). (b) A close-up view of the $\lambda/4$ IDC resonator. It is galvanically connected to the BS-SIF and shorted to ground. The IDC has a total of forty fingers extending from the centre conductor of the CPW, with two different lengths. (c) A lumped-element circuit model of the device. The resonator has two contributions to its total inductance: a kinetic inductance $L_k$ and a geometric inductance $L_0$. The former is non-linear and responsible for three-wave mixing, while the latter gives rise to an AC magnetic field $B_1$ that couples to \Bi\ donors implanted into the silicon substrate.} 
	\label{fig_SI:device_schematic}
\end{figure}

The design of the device is similar to that of several previously reported Kinetic Inductance Parametric Amplifiers (KIPAs) \cite{Parker2021,Vine2023}. It consists of two main components: a high-quality factor quarter-wavelength ($\lambda/4$) microwave resonator and a Band-Stop Stepped Impedance Filter (BS-SIF, Fig.~\ref{fig_SI:device_schematic}). The former is constructed as a dense Interdigitated Capacitor (IDC) with forty fingers of width $1~\mu$m extending from the center conductor, each of which has a spacing of $1.5~\mu$m to ground (Fig.~\ref{fig_SI:device_schematic}b). The fingers of the IDC were made to have two lengths (twenty with length $125~\mu$m and twenty with length $135~\mu$m) to produce a frequency detuning of the next harmonic at $\omega_1$. This ensures $\omega_1 \neq 3\omega_0$, so that a strong pump tone at $\omegap=2\omega_0$ does not couple the $\lambda/4$ and $3\lambda/4$ modes. The BS-SIF consists of eight total segments of CPW with alternating low $\Zlo=29.5~\Omega$ and high $\Zhi=123~\Omega$ impedance. The electrical lengths of each segment are $\lambda/4$ at the resonant frequency $\omega_0$, which results in a deep stop-band centered at $\omega_0$, and pass-bands at DC and $\omegap$. The BS-SIF therefore serves two purposes: it isolates the resonant mode from the measurement port (i.e. creates a large coupling quality factor $Q_c$) and it enables the resonator to be galvanically connected to the measurement port so that it can be biased with a DC current $\Idc$. The design parameters of the device are summarized in Table~\ref{table_SI:design_parameters}.

\begin{table}
\begin{center}
\begin{tabular}{|p{2.2cm}||p{2.2cm}|p{2.2cm}|p{2.5cm}| p{2.5cm}|}
\hline
 Parameter & $\Zhi$ (BS-SIF) & $\Zlo$ (BS-SIF) & Resonator Centre Conductor & Resonator Fingers \\
 \hline
 $W$~$(\mu$m) & 10 & 162 & 1 & 1 \\
 $G$~$(\mu$m)  & 82 & 6 & 2 & 1.5 \\
 Length~$(\mu$m) & 3218 & 3388 & 201 & 120, 130 \\
 $Z$~$(\Omega)$ & 123 & 29.5 & - & - \\
 \hline
\end{tabular}
\caption{\label{table_SI:design_parameters} KIPA design parameters. $W$ and $G$ are the width of the CPW conductor and gap, respectively. The impedance $Z$ of the BS-SIF segments is found through simulations in the software package Sonnet. The impedance of the IDC resonator is estimated from simulation to be $34~\Omega$ at $\omega_0$ and $33~\Omega$ at $\omegap$.}
\end{center}
\end{table}

\section{Measurements of Resonator Frequency and $Q$-Factor}

Measurements of $S_{11}$ were performed with a Vector Network Analyzer (VNA) (Keysight PNA-L N5231B). Due to the narrow bandwidth of the resonator relative to its frequency tunability, measurements of $S_{11}$ with the resonator far-detuned were subtracted to remove background ripple and correct for the line delay. For the lowest power measurements in Fig.~1d of the main text, the signal was digitally filtered with a Savitzky–Golay filter to improve the signal to noise ratio.

\begin{figure}
	\centering
	\includegraphics[width=\linewidth]{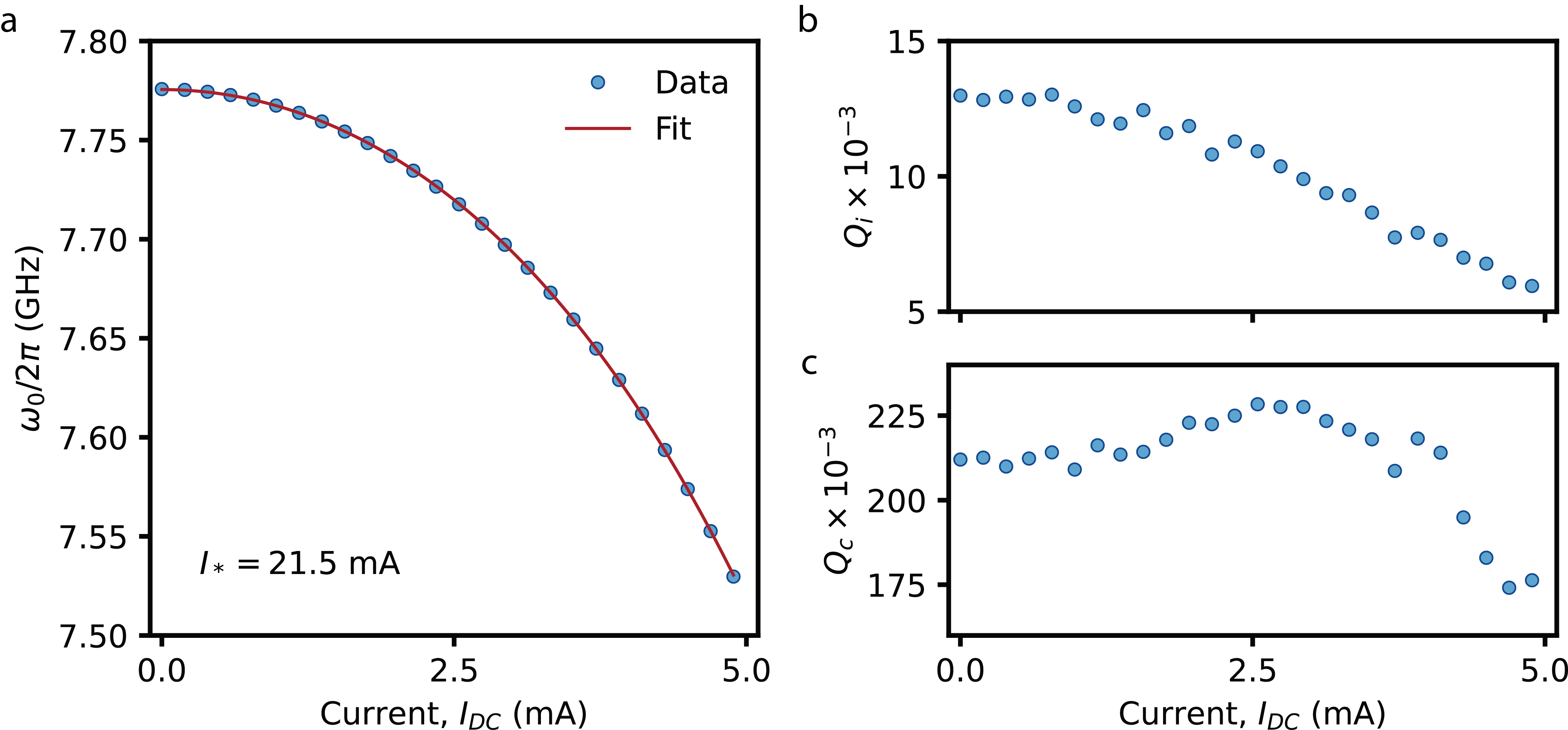}
	\caption{The resonance frequency and quality factors measured as a function of $\Idc$. The values are extracted for measurements of $S_{11}$ using a VNA with a power such that $\bar{n} \approx 1$. The data in (a) is fit with Eq.~\ref{eqn_SI:frequency_tunability} to extract the constant $I_*$.} 
	\label{fig_SI:Qs_vs_Idc}
\end{figure}

Fig.~\ref{fig_SI:Qs_vs_Idc} shows $\omega_0$, $Q_i$ and $Q_c$ measured as a function of $\Idc$, which are extracted from fits of $S_{11}$ to cavity input-output theory. The resonance frequency $\omega_0/2\pi$ could be tuned by $246~$MHz by applying $\Idc= 4.89~$mA (Fig.~\ref{fig_SI:Qs_vs_Idc}a). The kinetic inductance is expected to vary with $I_{DC}$ according to

\begin{equation}
    L_k(\Idc) = L_k(0)\left[1 + \frac{\Idc^2}{I^2_*} + \bigO(I^4)\right]
,\end{equation}

\noindent where $I_*$ is a constant \cite{Vissers2015}. This in turn results in the frequency of the resonator varying as \cite{Vissers2015,Parker2021}

\begin{equation}
    \omega_0(\Idc) = \omega_0(0)\left[1-\frac{\Idc^2}{2I^2_*}\right]
    \label{eqn_SI:frequency_tunability}
.\end{equation}

\noindent Fitting the measured $\omega_0(\Idc)$ with Eq.~\ref{eqn_SI:frequency_tunability} yields $I_*=21.5$~mA for this device. Notably, the device is extremely under-coupled over the entire range of operation, with $Q_i / Q_c$ in the range $[0.033,0.064]$ (Fig.~\ref{fig_SI:Qs_vs_Idc}b,c).

Fig.~1d of the main text shows measurements of $Q_i$ taken as a function of the applied microwave power $P_0$. This can be converted to an average number of intracavity photons $\bar{n}=2Q_L^2P_0/(\hbar\omega_0^2Q_c)$ where $Q_L=(Q_i^{-1}+Q_c^{-1})^{-1}$ is the loaded quality factor \cite{Bruno2015}. The general improvement of $Q_i$ with $\bar{n}$ is consistent with many other studies of high-$Q$ superconducting microwave resonators, where the influence of Two Level Systems (TLSs) on $Q_i$ has been well documented \cite{Wang2020a}. Notably, where TLSs limit $Q_i$, one typically expects $Q_i$ to flatten below $\bar{n}=1$, because the microwave power is insufficient to depolarize the TLSs. This is not observed in the range of signal powers studied here. We note that some authors have reported that $Q_i$ reduces with signal power down to $\bar{n}=10^{-3}$ \cite{Niepce2019a}.

\section{Model of a Parametrically-Pumped Duffing Oscillator}

To understand the behaviour of the KIPA when strongly pumped, we seek to develop a theoretical model based on a parametrically-pumped Duffing oscillator, which has previously proven to be an excellent description for Josephson Parametric Oscillators \cite{Wilson2010,Wustmann2013,Lin2014}. For a classical oscillator with position $x$, the Duffing equation is given by

\begin{equation}
    \ddot{x} + 2\lambda\dot{x} + \left(\omega_0^2 + F\cos(2\omega t)\right)x - dx^3 = 0
,\end{equation}

\noindent where $\lambda$ is the linewidth of the oscillator, $\omega$ is the frequency of a periodic drive with strength $F$, and $d$ is the Duffing constant. For a mass on a spring, the Duffing constant describes a softening or stiffening of the spring as it extends from its equilibrium position. We show below that for a superconducting microwave resonator it is related to the Kerr effect.

To relate the Duffing equation to the KIPA we take the approach of Lin, et al. \cite{Lin2014} and begin with the master equation for the intracavity field, which is given by

\begin{gather}
    \dot{a} = \frac{1}{i\hbar}[a,H_\mathrm{KIPA}] + \sqrt{\kappa}a_\mathrm{in}(t) + \sqrt{\gamma}b_\mathrm{in}(t) - \frac{\kappa + \gamma}{2}a 
    \label{eqn_master}
.\end{gather}

\noindent Here $a$ ($a^\dag$) is the bosonic annihilation (creation) operator for the cavity mode, $a_\mathrm{in}$ ($a^\dag_\mathrm{in}$) is the bosonic annihilation (creation) operator for the port mode, and $b_\mathrm{in}$ ($b^\dag_\mathrm{in}$) is the bosonic annihilation (creation) operator for the bath mode. $\kappa=\omega_0/Q_c$ and $\gamma=\omega_0/Q_i$ are the rates at which photons couple from the cavity mode to the port mode and bath mode, respectively.

The Hamiltonian for the KIPA was previously derived in Reference \cite{Parker2021}. In the frame rotating at half the pump frequency it is given by

\begin{gather}
    \frac{H_\mathrm{KIPA}}{\hbar} = \Delta a^\dag a + \frac{\zeta}{2}a^{\dag 2} + \frac{\zeta^*}{2}a^2 + \frac{K}{2}a^{\dag^2}a^2
    \label{eqn_H_kipa}
,\end{gather}

\noindent where

\begin{gather}
\Delta = (\omega_0 + \deltadc + \deltap + K - \frac{\omegap}{2})
\label{eqn_kipa_delta}
,\\
\deltadc = -\frac{1}{2}\frac{\Idc^2}{I_*^2}\omega_0
\label{eqn_kipa_current_response}
,\\
\deltap = -\frac{1}{8}\frac{\Ip^2}{I_*^2}\omega_0
\label{eqn_kipa_pump_response}
,\\
K = -\frac{3}{8}\frac{\hbar\omega_0}{L_T I_*^2}\omega_0
\label{eqn_kipa_kerr}
,\\
\zeta = -\frac{1}{4}\frac{\Idc\Ip}{I_*^2}\omega_0e^{-i\phip}
\label{eqn_kipa_zeta}
.\end{gather}

\noindent Here, $\Delta$ is a detuning of half of the pump frequency from the point of degeneracy (i.e. the cavity frequency), $\zeta$ is the three-wave mixing strength, and $K$ is the Kerr strength. $\Delta$ accounts for the shift of the cavity with an applied DC current ($\deltadc$) and pump current ($\deltap$) due to the nonlinear kinetic inductance in the KIPA. $I_*$ is a constant that sets the scale of the non-linearity of the KIPA and can be experimentally determined by measuring $\deltadc$ with a Vector Network Analyzer (VNA), as demonstrated in Fig.~\ref{fig_SI:Qs_vs_Idc}a. $\Idc$ is the applied DC current that enables three-wave mixing and $\Ip$ is the pump current peak amplitude. $L_T$ is the total kinetic inductance and $\phip$ is the phase of the pump tone.

Substituting Eq.~\ref{eqn_H_kipa} into Eq.~\ref{eqn_master} yields

\begin{equation}
 \dot{a} =-i\Delta a -i\zeta a^\dag - iK a^\dag a^2 - \frac{\kappa + \gamma}{2}a + \sqrt{\kappa}a_{\mathrm{in}}(t) + \sqrt{\gamma}b_{\mathrm{in}}(t)
 \label{eqn_master_2}
.\end{equation}

\noindent Next we consider the simplification $\braket{a_\mathrm{in}}=\braket{b_\mathrm{in}}=0$, which corresponds to the case where no resonant signals are sent to the cavity via the port or bath modes. In this case, Eq.~\ref{eqn_master_2} can be transformed into two real-valued coupled differential equations by substituting $a=X - iY$, where $X$ and $Y$ are the operators for the quadrature amplitudes of the intracavity field. Doing so yields

\begin{equation}
\begin{pmatrix} \dot{X} \\ \dot{Y} \end{pmatrix} = 
 \begin{pmatrix}
 -Y\Delta + Y\zeta - K(X^2+Y^2)B - \bar{\gamma}X 
,\\
 X\Delta + X\zeta + K(X^2+Y^2)X - \bar{\gamma}Y
 \end{pmatrix}
,\end{equation}

\noindent where $\bar{\gamma}=(\kappa + \gamma)/2$. This can be recast in the form 

\begin{equation}
\frac{1}{\bar{\gamma}}\frac{\partial}{\partial t}
\begin{pmatrix} X \\ Y \end{pmatrix} = 
 \begin{pmatrix}
 -X -\partial g(X,Y)/\partial Y\\
 -Y + \partial g(X,Y)/\partial X
 \end{pmatrix}
 \label{eqn_master_eqn_simplified}
,\end{equation}

\noindent where

\begin{equation}
g(X,Y) = \frac{\Delta}{2\bar{\gamma}}(X^2 + Y^2) + \frac{\zeta}{2\bar{\gamma}}(X^2 - Y^2) + \frac{K}{4\bar{\gamma}}(X^2 + Y^2)^2
\label{eqn_metapotential_kipa}
.\end{equation}

Previous authors have noted the similarity of Eq.~\ref{eqn_master_eqn_simplified} to Hamilton's equations \cite{Wilson2010,Wustmann2013}. This has lead to Eq.~\ref{eqn_metapotential_kipa} being described as a metapotential because the behaviour of the superconducting circuit is analogous to a particle traversing the potential surface $g(X,Y)$. 

In Eq.~\ref{eqn_metapotential_kipa}, the terms $\Delta/(2\bar{\gamma})$, $\zeta/(2\bar{\gamma})$, and $K/(2\bar{\gamma})$ correspond generalized versions of the detuning of the pump frequency, the pump strength, and the Duffing non-linearity, respectively. Their influence on the behaviour of the oscillator become clear by taking the steady state solution of Eq.~\ref{eqn_master_eqn_simplified}. We also make use of a change of variables for $X$ and $Y$ so that they can be described in terms of a single amplitude $\alpha$ and phase $\theta$ by making the substitutions $X=\alpha \cos(\theta)$ and $Y=\alpha \sin(\theta)$. Doing so yields

\begin{gather}
    \cot(\theta) = \frac{1}{\bar{\gamma}}\left(\zeta + \Delta + \alpha^2 K\right)
    \label{eqn_ss1}
    ,\\
    \tan(\theta) = \frac{1}{\bar{\gamma}}\left(\zeta - \Delta - \alpha^2 K\right)
    \label{eqn_ss2}
.\end{gather}

Equations~\ref{eqn_ss1} and \ref{eqn_ss2} are $\pi$-periodic, and thereby imply the existence of two non-trivial solutions. Moreover, it is straightforward to solve for $\alpha$ by taking their product, which yields a solution that is independent of $\theta$ and is equal to

\begin{equation}
\alpha = \bigg(\pm \frac{\sqrt{\zeta^2-\bar{\gamma}^2}-\Delta}{K}\bigg)^{1/2}
\label{eqn_so_boundary_kipa1}
.\end{equation}

Because $\alpha$ is an amplitude, we require Eq.~\ref{eqn_so_boundary_kipa1} to yield real solutions. This is true only when 

\begin{equation}
\zeta^2 > \Delta^2 + \bar{\gamma}^2
\label{eqn_so_boundary_kipa2}
,\end{equation}

\noindent so that we can interpret Eq.~\ref{eqn_so_boundary_kipa2} as a boundary in parameter space. When the inequality is false, the device functions as a linear parametric amplifier. In this case, provided $\braket{a_\mathrm{in}}=0$, $\alpha$ will remain equal to zero. When the inequality is true, however, the resonator will quickly develop a large intracavity field with amplitude $\alpha$ and phase $\theta = 0$ or $\theta=1\pi$.

\section{Hysteretic Parametric Self-Oscillations}

\begin{figure}
	\centering
	\includegraphics[width=\linewidth]{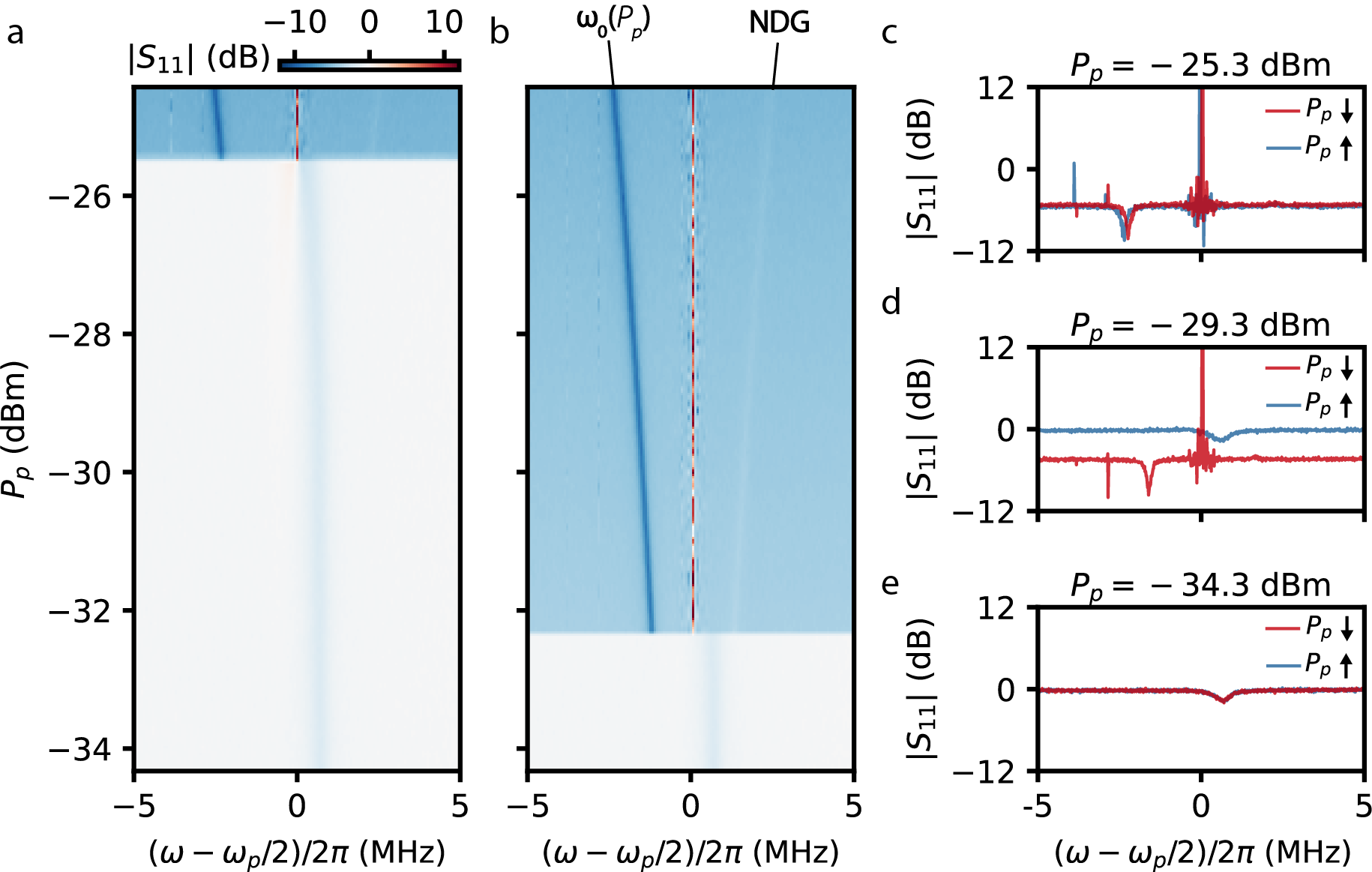}
	\caption{Hysteresis of self-oscillations. (a) VNA measurements of $|S_{11}|$ as $\Pp$ is increased. (b) $|S_{11}|$ as $\Pp$ is decreased. The labels $\omega_0(\Pp)$ and NDG show the shifted resonance frequency and the idler associated with a non-degenerate gain process, respectively. (c-e) Line cuts from (a) and (b) at various powers. The DC current was set to $\Idc = 0.83$~mA for these measurements. The sharp change in the baseline power is due to the saturation of the amplification chain when the device self-oscillates. The colourmap for (a) was truncated at +12 dB and also applies to panel (b).} 
	\label{fig_SI:self_osc_VNA}
\end{figure}

In Fig.~2 of the main text it is demonstrated that the detector signal latches. This is related to the fact that $Q_i$ is dependent upon $\bar{n}$, which results in hysteresis of $\Pth$. This hysteresis can be directly observed when measuring $|S_{11}|$ with a VNA while sweeping $\Pp$. In Fig.~\ref{fig_SI:self_osc_VNA}a we increase $\Pp$ and observe a sharp transition in the behaviour of the device as $\Pp$ is raised beyond $-25.5~$dBm. In contrast, when $\Pp$ is decreased, the behaviour changes at $\Pp=-32.4~$dBm (Fig.~\ref{fig_SI:self_osc_VNA}b).

Line cuts taken from the two measurements show that the device functions as a simple resonator for $\Pp < -32.3$~dBm (Fig.~\ref{fig_SI:self_osc_VNA}e) and as a parametric oscillator for $\Pp > -25.4$~dBm (Fig.~\ref{fig_SI:self_osc_VNA}c). The behaviour observed at large $\Pp$ can be clearly identified as parametric self-oscillations due to the large power generated at exactly half the pump frequency. For $\Pp$ intermediate to these values, the device functions as a parametric amplifier prior to latching. Notably, the linear parametric gain achieved is much smaller than in previous studies using KIPAs \cite{Parker2021,Vine2023}, which is a consequence of the fact that $Q_i < Q_c$ for this device. The power generated in the self-oscillating state is large enough to cause the amplification chain to compress, which results in the baseline shifting by -5.6~dB. For a parametric oscillator, one would also expect that the amplitude of the self-oscillations would be arrested by a non-linear shift in the resonance frequency due to the Duffing non-linearity. This is because for a fixed $\omegap$ and $\Pp$, shifting $\omega_0$ will reduce the rate of down-conversion, which in combination with a finite cavity bandwidth set by $Q_L$, eventually leads the intracavity field to a high-amplitude equilibrium. Fig.~\ref{fig_SI:self_osc_VNA}b clearly reveals this behaviour, with the detuning $(\omegap/2 - \omega_0(\Pp))/2\pi$ reaching a maximum of $3.2$~MHz at the highest $\Pp$ measured.

It is also notable that the depth of the resonance appears to increase when the device self-oscillates. Fitting the resonance in isolation from the other features confirms this and reveals that it is primarily the result of $Q_i$, which increases from $13\times 10^3$ to $50\times 10^3$ when the device self-oscillates. We attribute this to be the result of the down-converted power partially saturating the TLSs.

\section{Evidence of Additional Mixing Processes}

\begin{figure}
	\centering
	\includegraphics[width=89mm]{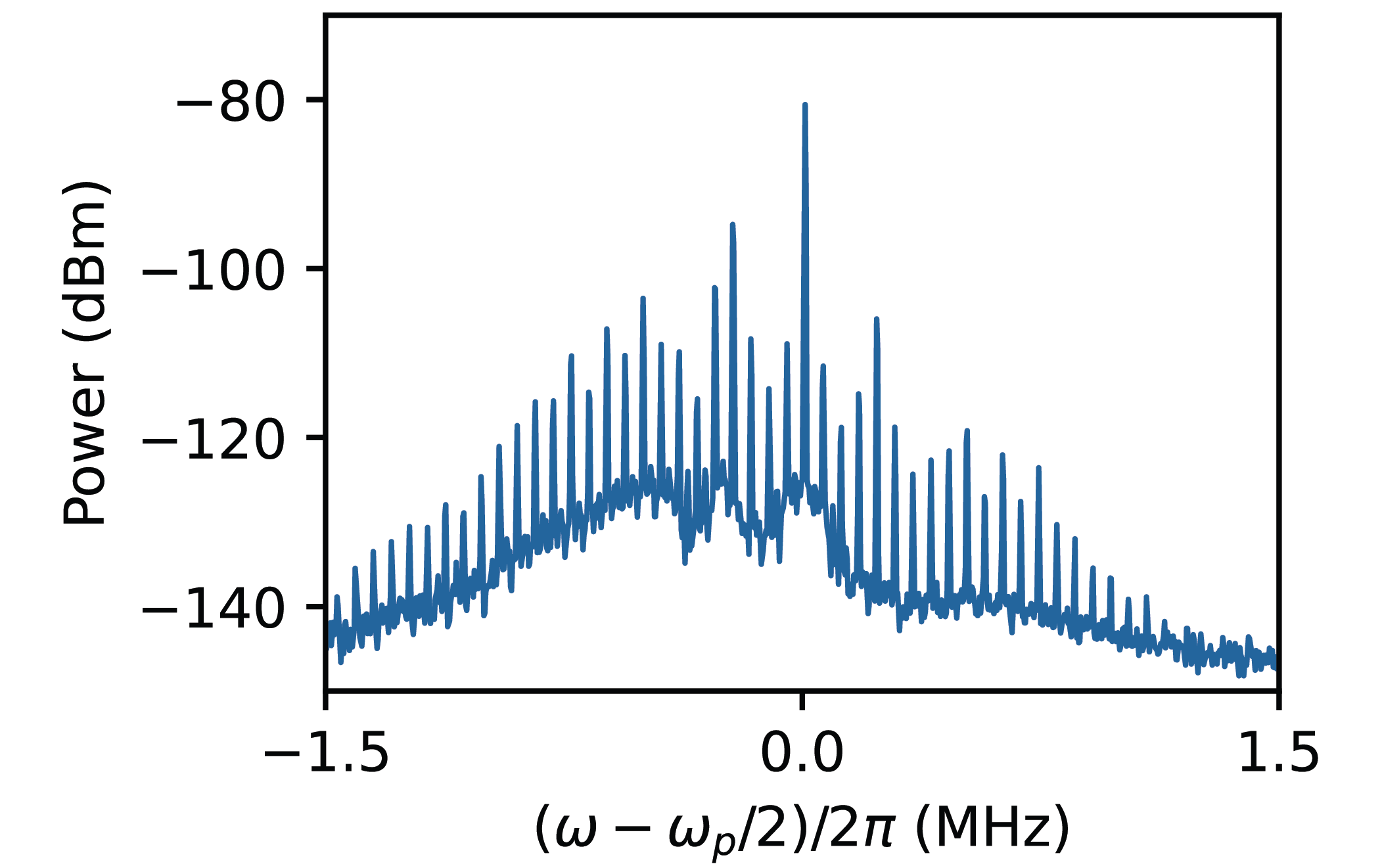}
	\caption{A spectrum of the self-oscillation signal. For this measurement $\Idc=2.5~$mA and the resonator frequency was measured to be $\omega_0/2\pi=7.7203~$GHz. The only signals supplied to the device during the measurement are the DC current $\Idc$ and a pump with frequency $\omega_p=2\omega_0$ and power $\Pp=-50.3~$dBm. All powers are referred to the device input.}
	\label{fig_SI:SO_spectrum}
\end{figure}

The measurements of $S_{11}$ shown in Fig.~\ref{fig_SI:self_osc_VNA} reveal several features that are not expected from the basic theory of a parametric oscillator, namely several sharp peaks on the red sideband and fine structure near $\omega=\omegap/2$. To better understand these features, we measure the output signal with a spectrum analyser centered on $\omega_0$ (as measured with a VNA with the pump off) when the device is supplied only a DC current and a pump with frequency $\omega_p=2\omega_0$. Fig.~\ref{fig_SI:SO_spectrum} shows one spectrum obtained with $\Idc=2.5~$mA and $\Pp=-50.3~$dBm. For this setpoint, $\Pp$ is just above the level required to initiate parametric self-oscillations, but similar results are obtained for larger $\Pp$ and for different $\Idc$. The spectrum reveals a frequency comb that spans nearly the entire $3~$MHz bandwidth of the measurement, with teeth that are equally spaced by 54~kHz. The tooth of the comb with the largest power is indeed centered at $\omega_p/2$, as expected for a parametric oscillator. The generation of frequency combs with superconducting microwave resonators has been demonstrated in several works including References \cite{Erickson2014,Cassidy2017a,Khan2018,Lu2021,Wang2021a} with several mechanisms underlying comb formation. We also note that the spectrum is obtained with phase modulation of the pump disabled, and direct measurements of the microwave sources do not reveal any sidebands that might otherwise explain the frequency comb generated by the KIPO. While a full explanation of the comb generated by the KIPO is beyond the scope of this work, it nevertheless has important implications to the experiments of this manuscript. First, when the KIPO is made to self-oscillate, the combs introduce a time-varying component to the amplitude of the signal demodulated with a local oscillator of frequency $\omega_\mathrm{LO}=\omega_p/2$, as in Figs.~2b and 4b of the main text. Second, the equal spacing of the teeth suggests they are likely generated via higher-order mixing processes (e.g. four wave mixing), which would thereby deplete power from the tooth at $\omega_p/2$. As the frequency comb is not captured in the model of a Duffing oscillator used to describe the KIPO, the amplitude of the self-oscillating state given by the model (Eq.~\ref{eqn_so_boundary_kipa1}) will not match that of the experiments. Because the frequency comb is generated only upon the initiation of parametric self-oscillations, however, the Duffing oscillator model nevertheless remains applicable for estimating the parametric self-oscillation threshold $\Pth(\Pp,\omega_p)$.

\section{Mapping the Self-Oscillation Boundary}

In this section we show how $\Pth$, which is written in terms of the KIPA Hamiltonian in Eq.~\ref{eqn_so_boundary_kipa2}, can be compared to experiments.

We begin by expanding the $\Delta^2$ term using Eq.~\ref{eqn_kipa_delta}, which yields

\begin{equation}
\begin{split}
\Delta^2 & = \omega_0^2 + \deltadc^2 + \deltap^2 + K^2 + \frac{\omegap^2}{4} \\
 & + 2\omega_0\left(\deltadc + \deltap + K - \frac{\omegap}{2}\right) \\
 & + 2\deltadc\left(\deltap + K - \frac{\omegap}{2}\right) \\
 & + 2\deltap\left(K - \frac{\omegap}{2}\right) - K\omegap.
\end{split}
\label{eqn_deltasq_full}
\end{equation}

To simplify this we make note of the relative scales of each term. $\omega_0/2\pi$ and $\omegap/2\pi$ are both of order GHz. For our experiments, $\deltadc/2\pi$ is tens of MHz. The expression for $\deltap$ is given by Eq.~\ref{eqn_kipa_pump_response} where it is written in terms of $\Ip^2$. To get a sense of its magnitude we can write it in terms of a microwave power $\Pp$ by substituting $\Ip^2 \rightarrow 2\Pp/\Zp$, where $\Zp$ is the impedance of the device at frequency $\omegap$. From simulations of the device (Sonnet) we expect $\Zp\approx 33~\Omega$. For $\Pp=-33~$dBm, the largest power used for the experiment shown in Fig.~1c of the main text, this yields $\deltap/2\pi \approx 64~$kHz. From previous works we know that $K/2\pi$ is of order Hz for these devices \cite{Parker2021}. We therefore conclude that $\omega_0,\omegap \gg \deltadc \gg \deltap \gg K$ which allows us to approximate Eq.~\ref{eqn_deltasq_full} as

\begin{gather}
\begin{split}
\Delta^2  & \approx \underbrace{\omega_0^2 + \frac{\omegap^2}{4} - \omega_0\omegap}_{\GHz^2} + \underbrace{2\omega_0\deltadc - \omegap\deltadc}_{\GHz\times\MHz} + \underbrace{2\omega_0\deltap - \omegap\deltap}_{\GHz\times\kHz} + \underbrace{\deltadc^2}_{\MHz^2}
\end{split}
\label{eqn_deltasq_approx}
,\end{gather}

\noindent where we have omitted the terms which are lower in frequency.

Here we note that in the main text we define $\Deltap=\omegap - 2\omega_0(\Idc)$, where $\omega_0(\Idc)$ is an experimentally measured value of the resonant frequency with an $\Idc$ applied. Written in terms of Eqs.~\ref{eqn_kipa_current_response}-\ref{eqn_kipa_kerr} this is

\begin{equation}
\Deltap = \omegap - 2\underbrace{(\omega_0 + \deltadc)}_{\omega_0(\Idc)} - \underbrace{2(\deltap + K)}_{\approx 0} \approx \omegap - 2(\omega_0 + \deltadc)
,\end{equation}

\noindent where we neglect the small shift in the resonance frequency due to the microwave power of the tone used to measure $S_{11}$. This is an important difference given that $\Delta$ accounts for the shift $\deltap$ due to the pump power, whereas $\Deltap$ does not. Using the approximation in Eq.~\ref{eqn_deltasq_approx} we find

\begin{gather}
\Delta^2 \approx \left(\frac{\Deltap}{2}\right)^2 + \deltap(2\omega_0-\omegap)
\\
= \left(\frac{\Deltap}{2}\right)^2 - \deltap(\Deltap + 2\deltadc)
\label{eqn_deltasq_subversion}
.\end{gather}

\noindent Here we see that Eq.~\ref{eqn_deltasq_subversion} contains a term that is proportional to $\deltap$ that accounts for the shift of the resonant frequency with the pump power.

Using Eqs.~\ref{eqn_kipa_delta}-\ref{eqn_kipa_zeta} and Eq.~\ref{eqn_deltasq_subversion}, Eq.~\ref{eqn_so_boundary_kipa2} can be written as

\begin{gather}
\zeta^2 > \left(\frac{\Deltap}{2}\right)^2 - \deltap(\Deltap + 2\deltadc) + \bar{\gamma}^2
\\
\frac{1}{16}\frac{\Idc^2\Ip^2}{I_*^4}\omega_0^2e^{-2i\phip} > \left(\frac{\Deltap}{2}\right)^2 +\frac{1}{8}\frac{\Ip^2}{I_*^2}\omega_0\left(\Deltap - \frac{\Idc^2}{I_*^2}\omega_0\right) + \frac{\omega_0^2}{4}(Q_i^{-1} + Q_c^{-1})^2
.\end{gather}

\noindent We then solve for $\Ip^2$ and convert it to a pump power as above. Without loss of generality, we assume $\zeta$ to be real and a positive value, i.e. we ignore the phase $\phi_p$, which is only relevant when a coherent signal is being amplified, whereas here we consider only vacuum noise. We also substitute $\Zp \rightarrow \alphap \Zp$, where $\alphap$ is a constant that accounts for ripple in the transmission of the pump power through the device. This yields

\begin{equation}
\Pp > \Pth = \frac{\alphap\Zp}{2}\frac{\left(\frac{\Deltap}{2}\right)^2 + \frac{\omega_0^2}{4}(Q_i^{-1} + Q_c^{-1})^2}{\frac{3}{16}\frac{\Idc^2\omega_0^2}{I_*^4} - \frac{1}{8}\frac{\omega_0\Deltap}{I_*^2}}
.\end{equation}

For the models shown in Fig.~1c of the main text we set $Q_c=220\times 10^3$ and $I_*=21.5~$mA based on the measurements shown in Fig.~\ref{fig_SI:Qs_vs_Idc}. We set $\Zp=33~\Omega$ based on simulations of the device (Sonnet) and $\alphap=1.3$ (1.13~dB) because it reproduces the experimental data well (the width of the self-oscillating region is primarily set by $\alphap \Zp$). With other KIPAs we have observed that $\alphap$ can vary by as much as $10$~dB over the operating frequency range (the ripple can can be inferred from measurements of the $\Pp$ required to achieve a fixed gain while tuning $\omega_0(\Idc)$). We also note that in Fig.~1c we apply an offset of $\Deltap/2\pi=-500$~kHz to the x-axis to center the parameter region where we observe parametric self-oscillations on $\Deltap=0$; this likely indicates that the frequency of the device drifted between the times Fig.~1c and $\omega_0(\Idc)$ were measured.

Overall, the model qualitatively reproduces the self-oscillation boundary that is measured. The main effect of increasing $Q_i$ is to shift the boundary to lower $\Pp$. Good agreement between the model and the data occurs when $Q_i=18\times 10^3$. This is notable because $Q_i(\bar{n}=0.01)\approx 5\times 10^3$ is measured when the pump is off (Fig.~1d of the main text). This might indicate that our model is incomplete, or that $Q_i$ is not independent of $\Pp$, as we have assumed thus far. Supporting the latter hypothesis is the recent work by Qiu, et al. who found that it was essential to consider the $\Pp$-dependent saturation of the TLSs due to amplified vacuum noise to quantitatively explain the behaviour of their amplifier \cite{Qiu2022}.

The main takeaway we highlight is that the model gives reasonable estimates for the observed $\Pth$. Moreover, because the device is undercoupled, its linewidth, and hence $\Pth$, is very sensitive to $\bar{n}$. This is the basis of our detector: $\Pth$ can be dynamically reduced below a fixed $\Pp$ when the device absorbs resonant power, resulting in onset of parametric self-oscillations.

\section{Phase-Coherent Measurements Resolving the Quiet, $0\pi$, and $1\pi$ States}

\begin{figure}
	\centering
	\includegraphics[width=\linewidth]{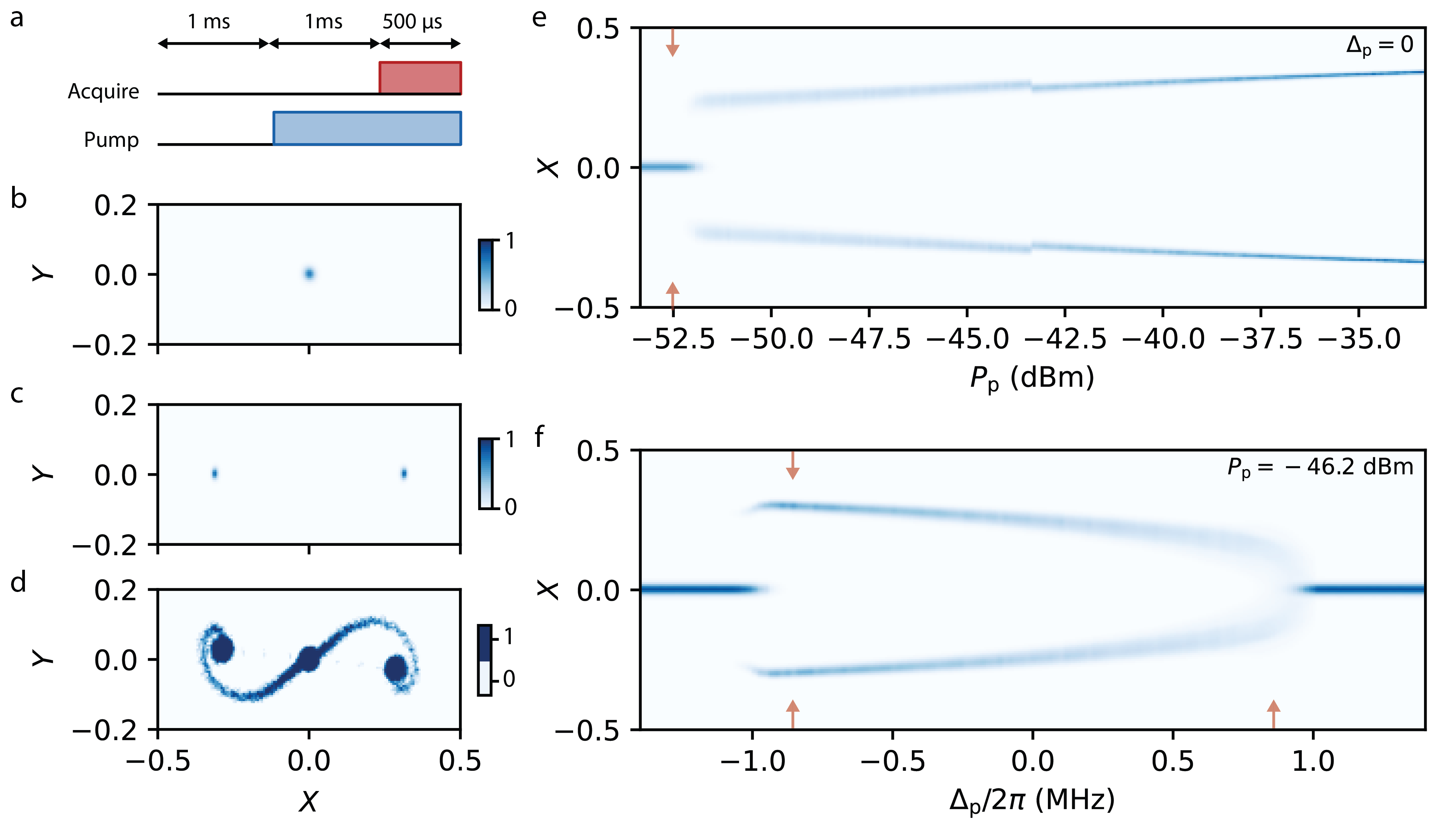}
	\caption{Histograms of demodulated self-oscillations. (a) The pulse sequence used to measure the histograms. It is repeated 500 times for each measurement. (b) A histogram where $\Pp < \Pth$, showing only the quiet state. (c) A histogram where $\Pp > \Pth$, showing the $0\pi$ and $1\pi$ states. (d) A histogram where $\Pp \approx \Pth$, showing the transition from the quiet to the self-oscillating states. Note that for this panel the histogram values are binary so the trajectories between the states can be easily seen. (e) A series of histograms in the $XY$-plane that have been projected onto $X$, showing the population evolve from the quiet state to the $0\pi$ and $1\pi$ states as $\Pp$ is increased. The red arrows correspond to $\Pth$ predicted from our model. (f) The same as in (e), except $\Deltap$ is varied for a fixed $\Pp=-46.2$~dBm. The red arrows correspond to the $\Deltap$ at which the model predicts the onset of parametric self-oscillations.}
	\label{fig_SI:SO_histograms}
\end{figure}

So far, our measurements have shown that when $\Pp > \Pth$ the device generates a high-amplitude field at frequency $\omegap/2$. Central to the model of a parametrically driven Duffing oscillator, however, is that there are two phase-coherent self-oscillating states, referred to as the $0\pi$ and $1\pi$ states.

To directly resolve these states, we down-convert the signal emitted from the device using a mixer driven by a local oscillator with frequency $\omega_\mathrm{LO}=\omegap/2$ and digitize the resulting quadrature amplitudes, $X(t)$ and $Y(t)$. In Fig.~\ref{fig_SI:SO_histograms}b we show a histogram of the signal in the $XY$-plane when the device is pumped with $\Pp < \Pth$. This is the quiet state of a parametric oscillator and corresponds to amplified vacuum noise; it is therefore a circle that rests at the origin of the $XY$-plane. In Fig.~\ref{fig_SI:SO_histograms}c we set $\Pp>\Pth$ and observe the emergence of two high-amplitude phase-coherent states. These are the $0\pi$ and $1\pi$ states. Their phase with respect to one another is fixed and equal to $\pi$, but their specific orientation in the $XY$-plane is set by their phase with respect to the local oscillator. We therefore choose to align both states along the $Y$ axis during post-processing. We note that to observe both the $0\pi$ and $1\pi$ states requires measuring the system many times. The histograms in Fig.~\ref{fig_SI:SO_histograms}b,c correspond to 250~ms of digitized data, which was acquired from 500~shots of the pulse sequence depicted in Fig.~\ref{fig_SI:SO_histograms}a. The pulse sequence provides 1~ms of dead time to ensure the device resets, followed by 1~ms where the pump is turned on, and then 500~$\mu$s where the pump remains on and the demodulated signal is digitized. The 1~ms period where the pump is turned on prior to digitizing the signal is intended to allow the device to reach its steady state. Nevertheless, when $\Pp\approx \Pth$, occasionally the device is seen to transition from the quiet state to either the $0\pi$ or $1\pi$ states. These transitions are effectively irreversible while the pump remains on, due to the hysteresis observed in Fig.~\ref{fig_SI:self_osc_VNA}. These transitions can be seen clearly by renormalizing the colormap so that every bin is either zero (no counts) or one (one or more counts), which show that the trajectories taken from the quiet to the $0\pi$ and $1\pi$ states are coherent and reproducible (Fig.~\ref{fig_SI:SO_histograms}d).

It is important to ensure no resonant power is supplied to the device during these measurements. It is well known that even a weak tone can cause the device to favour the $0\pi$ or $1\pi$ states, a phenomenon commonly referred to as injection locking \cite{Lin2014}. In the present experiments, the balanced weighting of the $0\pi$ and $1\pi$ states confirms injection locking is not occurring.

By acquiring a series of these histograms, we can map the time-averaged populations of the quiet, $0\pi$, and $1\pi$ states as a function of experimental parameters. In Figs.~\ref{fig_SI:SO_histograms}e,f we show series of histograms measured as a function of $\Pp$ and $\Deltap$, respectively. These measurements are equivalent to taking vertical and horizontal line-cuts across the self-oscillation boundary, which is mapped in Fig.~1c of the main text. For both sets of measurements, the self-oscillation boundary as predicted by the model described above is indicated by red arrows and shows good agreement with the measurements. The device parameters used in the model are the same as in Fig.~1c, with $Q_i=18\times 10^3$. Note that as we did for Fig.~1c, we apply here an offset of $\Delta_p=-400~$kHz to account for drift in the resonance frequency.

While the amplitude of the self-oscillating states increases with $\Pp$ and decreases with $\Deltap$, as broadly predicted by the model in Eq.~\ref{eqn_so_boundary_kipa1}, we note that the model does not produce a quantitative match to this aspect of our experiments. As stated above, this is because the model does not account for the emergence of the frequency comb shown in Fig.~\ref{fig_SI:SO_spectrum} which deplete the power at $\omegap/2$. Future refinements to the model, such as the inclusion of additional mixing processes and a $\Pp$-dependent $Q_i$, may facilitate a quantitative match to all aspects of our experiments.

\section{Detector Dark Count Rate}

\begin{figure}
	\centering
	\includegraphics[width=89mm]{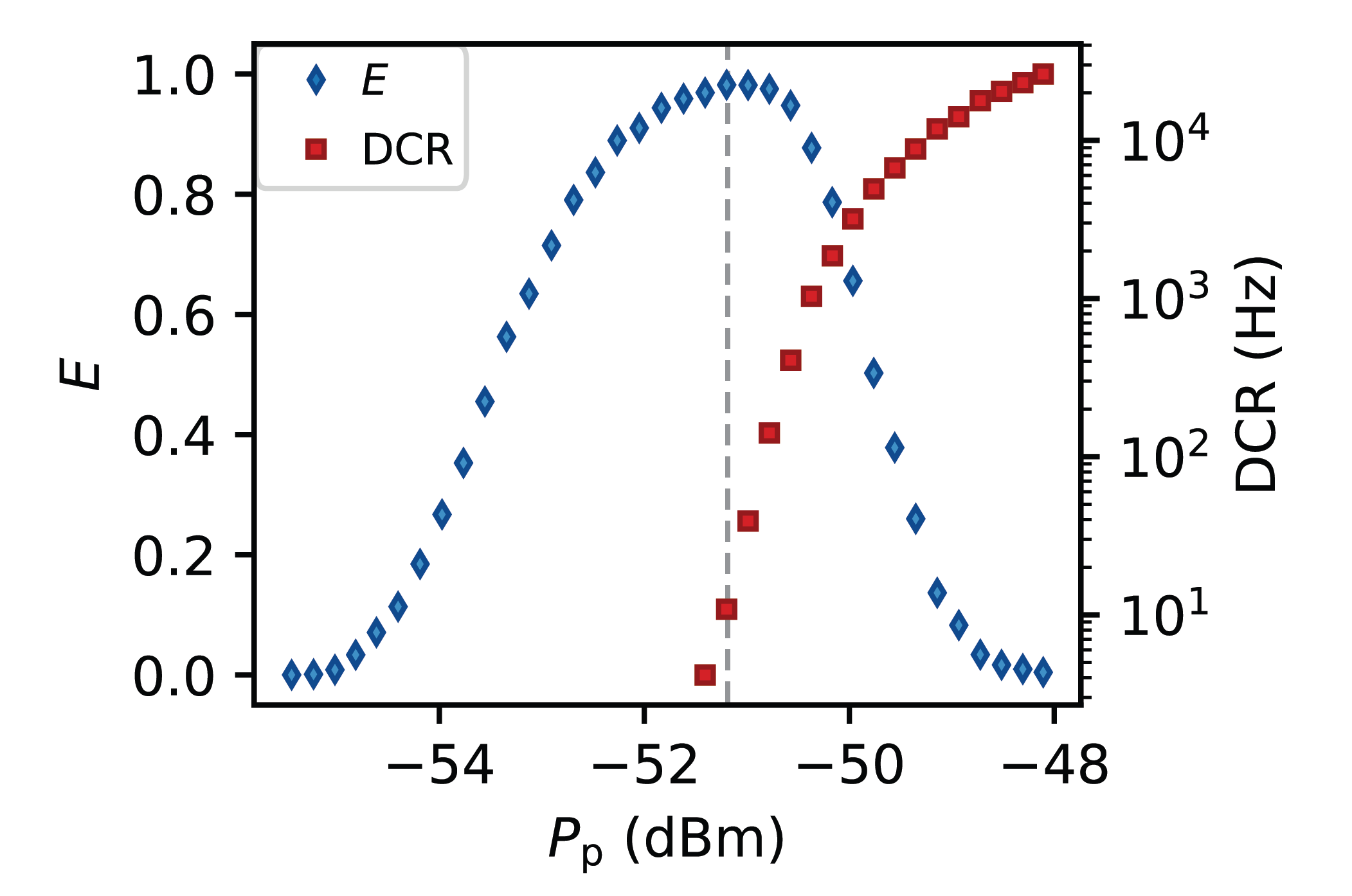}
	\caption{KIPO dark count rate. A comparison between the detection efficiency $E$ and DCR. The data is from the same experiment as shown in Fig.~3c of the main text, where $E$ was measured for pulses with duration $\tau_1=10~\mu$s and power $P_0=-111~$dBm. The vertical dashed line corresponds to $\Pp=-51.2~$dBm, which is the $\Pp$ used in Fig.~3e-g of the main text.}
	\label{fig_SI:dark_count_rate}
\end{figure}

In Fig.~3d of the main text we show the probability of measuring dark counts $\Pdark$ as a function of the pump power $\Pp$. From this measurement we also determine the dark count rate (DCR). To do so, we make use of the fact that the detector latches in the self-oscillating state. We can then calculate the DCR as

\begin{equation}
    \text{DCR} = \frac{n}{N \tau_\mathrm{tot} - \sum_i^n (\tau_\mathrm{tot} - t_i)}
\end{equation}

\noindent where $\tau_\mathrm{tot}=\tau_0 + \tau_1 + \tau_2 = 120~\mu$s is the total duration of the pump pulse in each shot of the control experiment (Fig.~3b of the main text), $N=10^4$ is the total number of shots of the control pulse sequence (depicted in Fig.~3b of the main text), $n$ is the number of dark counts, and $t_i$ is the time at which the detector ``clicks'' measured from the rising edge of the pump pulse.

In Fig.~\ref{fig_SI:dark_count_rate} we plot the DCR as a function of $\Pp$. We also directly compare it to the detector efficiency $E$ measured during the same experiment with pulses of duration $\tau_1=10~\mu$s and power $P_0=-111~$dBm. The minimum dark count rate we can measure is limited by the total measurement time and is equal to $1/(N\tau_\mathrm{tot})=0.83$~Hz. For $\Pp < -51.4~$dBm we detect no dark counts, whereas for $\Pp > -50.4~$dBm the DCR exceeds $1$~kHz. 

In Figs.~3e-g of the main text we determine the detection sensitivity and demonstrate that $E$ is phase-sensitive. For these experiments we set $\Pp=-51.2~$dBm, where in Fig.~\ref{fig_SI:dark_count_rate} we measure a total of $n=13$ dark counts in $N=10^4$ shots, corresponding to a DCR of $10.8~$Hz.

\section{Receiver Operating Characteristic Curve}

\begin{figure}
	\centering
	\includegraphics[width=89mm]{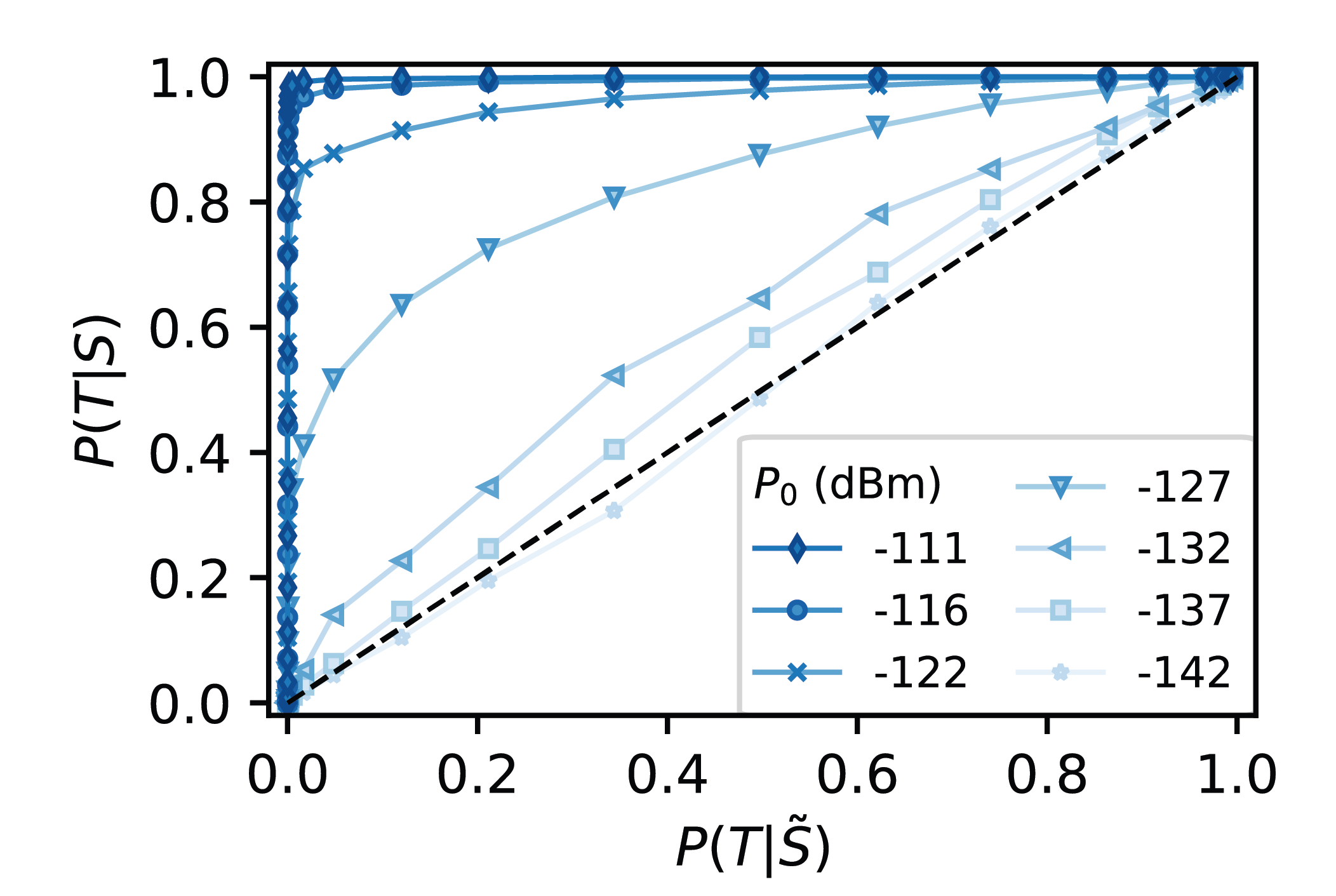}
	\caption{Receiver operating characteristic curve. The data is the same as is shown in Fig.~3c of the main text. A black diagonal line is plotted along the diagonal $\Pbright=\Pdark$ and corresponds to a random binary classifier.}
	\label{fig_SI:ROC_curve}
\end{figure}

There are a wide variety of metrics that can be used to benchmark the performance of a detector. Throughout the main text we use the detector efficiency $E=\Pbright-\Pdark$ as a single metric. One disadvantage of this approach, however, is that by itself $E$ does not communicate the specific balance between the probability of successful detection $\Pbright$ and the probability of dark counts $\Pdark$, both of which may vary with different settings of the detector. A common approach to evaluating these trade-offs is to plot an receiver operating characteristic (ROC) curve, which compares $\Pbright$ directly to $\Pdark$. In Fig.~\ref{fig_SI:ROC_curve} we show an ROC curve for the experiment depicted in Fig.~3c of the main text. The black diagonal line corresponds to the scenario where $\Pbright=\Pdark$, in which case the detector behaves as a random binary classifier, i.e. using the detector is equivalent to guessing. For the pulses with power $P_0=-137~$dBm, the detector outperforms the random binary classifier. For this experiment, the pulses had duration $\tau_1=10~\mu$s, so that the total energy within the wavepacket was $J=P_0\tau_1=0.21_{0.17}^{0.27}~$zJ ($42_{33}^{52}$ photons), where the upper and lower values correspond to a $1$~dB uncertainty in $P_0$.

\section{Detector Efficiency with Bias Current}

\begin{figure}
	\centering
	\includegraphics[width=\linewidth]{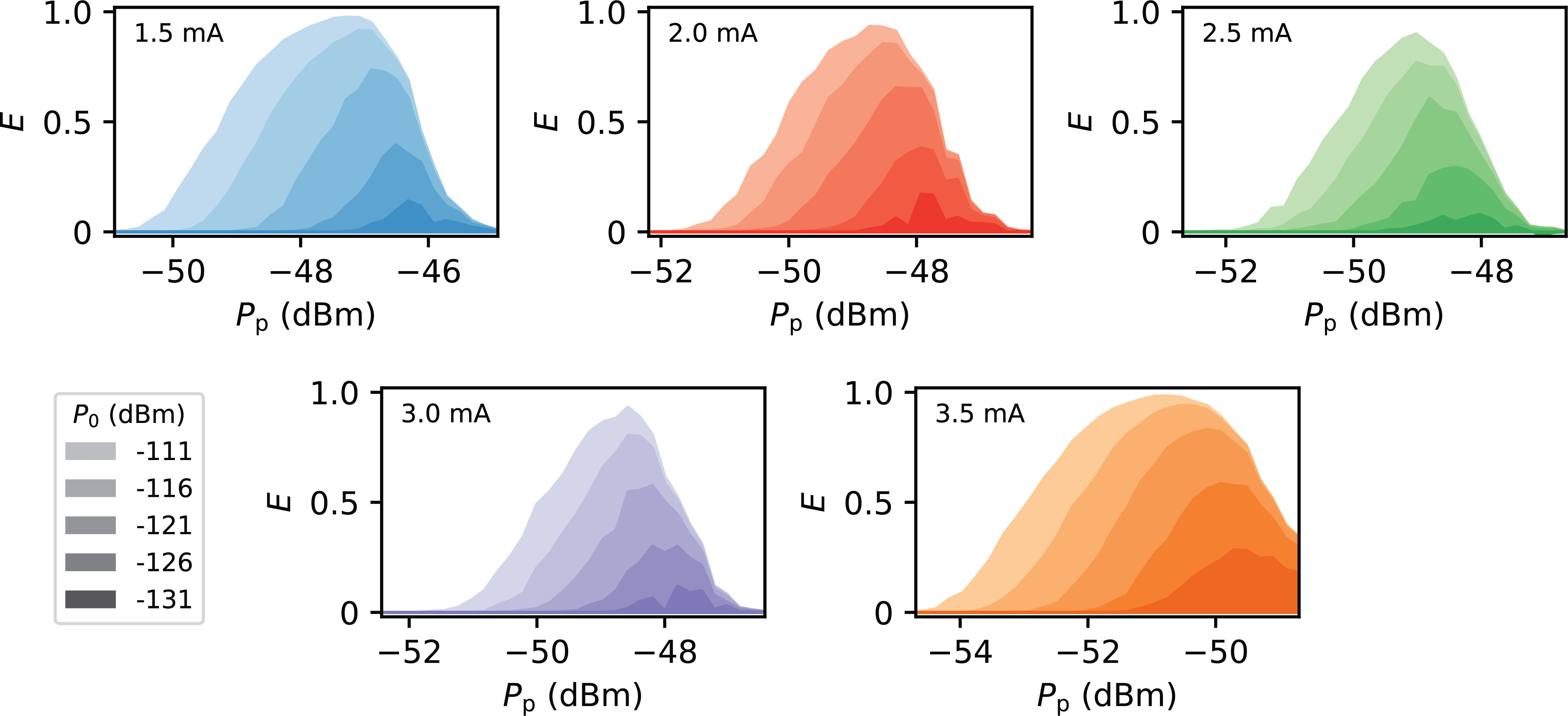}
	\caption{The $E$ measured as a function of $\Pp$ for a range of $\Idc$ setpoints. Each datapoint corresponds to $5\times 10^3$ shots of both pulse sequences depicted in Figs.~3a,b of the main text, with $\tau_1=10~\mu$s. The legend in the lower left corner applies to all panels.}
	\label{fig_SI:E_vs_Idc}
\end{figure}

In Fig.~\ref{fig_SI:E_vs_Idc} we compare measurements of $E$ completed over a series of $\Idc$ setpoints, corresponding to a tunable frequency range of $\deltadc/2\pi \approx 95~$MHz. For each setpoint, the device achieves $E$ near unity for pulses with duration $\tau_1=10~\mu$s and $P_0=-111~$dBm. The similar performance of the device at each setpoint indicates it is generally insensitive to $\Idc$, and can therefore be used as a detector over much of its total tunable frequency range.

\section{Measuring ESR as a Function of Magnetic Field}

\begin{figure}
	\centering
	\includegraphics[width=\linewidth]{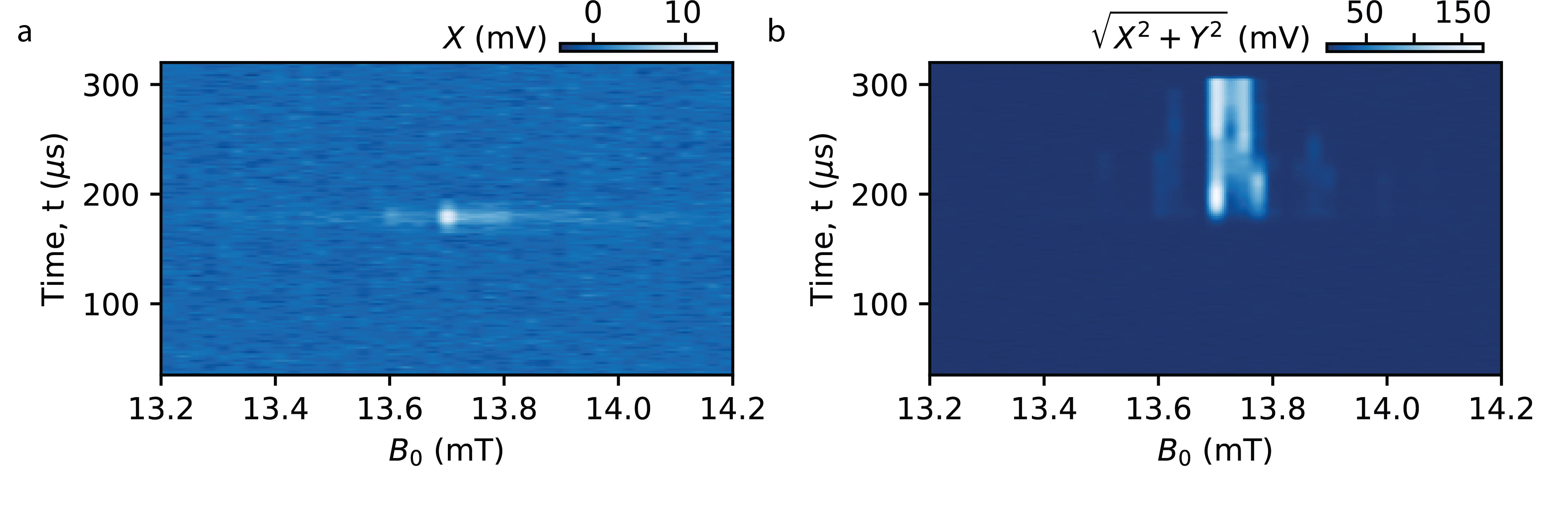}
	\caption{Spectroscopy with CPMG measurements. (a) The amplitude of a signal measured with the conventional CPMG measurement sequence. (b) The same as (a) but measured with the modified CPMG sequence, with $\Pp = -40.0$~dBm. In both measurements we perform twenty shots of the CPMG sequence with $N=20$ refocusing pulses and average the entire record. We recorded no dark counts for the control sequence of the measurement in (b). The measurements in both panels were taken during the same field sweep, with $\Idc = 2.53$~mA and a waiting period of $7~$s between each shot of each pulse sequence.}
	\label{fig_SI:field_sweep}
\end{figure}

To simulate the use of the KIPO as a spectroscopic probe, we measure ESR as a function of the magnetic field strength $B_0$ for the $\bra{-,4}S_x\ket{+,5}$ transition of {\Bi}. In Fig.~\ref{fig_SI:field_sweep}a we show a conventional ESR measurement using a CPMG sequence with $N=20$ refocusing pulses. For each $B_0$ we execute the full pulse sequence a total of twenty times, and average the full record of $20\times 20 = 400$ echoes. We show the signal measured on $X$, the quadrature onto which the spin echo signal was emitted, and find the spin signal to have a maximum at $13.7~$mT. 

We compare the above to a measurement that is otherwise equivalent, except that the KIPO is used as a ``click''-detector by biasing it with a pump with power $\Pp=-40.0$~dBm during the time period the echoes are refocused (Fig.~\ref{fig_SI:field_sweep}b). The self-oscillation signal that is triggered by the spin echoes cannot be aligned onto a single quadrature, so we instead plot the amplitude of the full homodyne-demodulated signal $\sqrt{X^2 + Y^2}$. We show the average signal acquired over all $20\times 20=400$ refocusing pulses which is justified for this measurement because no dark counts were recorded for the control experiment (the sequence omitting the $x_{\pi/2}$ pulse was used). As for the conventional measurement, the spectrum shows a maximum at $13.7~$mT, indicating that the detector can be used to determine the $B_0$ (frequency) of the ESR transition. Moreover, the amplitude of the detector signal with respect to the noise of the measurement is far greater than in the conventional case. This is despite the fact that we purposefully operate the detector at a point of low detection efficiency to avoid any dark counts. We refrain from quantitatively comparing the signal to noise ratio (SNR) of these measurements as has been done previously for experiments using linear amplifiers because the detection signals are fundamentally different; because the detector signal latches, it can be used to achieve an arbitrary SNR by extending the time the signal is integrated.

\section{Histograms of Detector ``Clicks'' in a CPMG ESR Measurement}

\begin{figure}
	\centering
	\includegraphics[width=\linewidth]{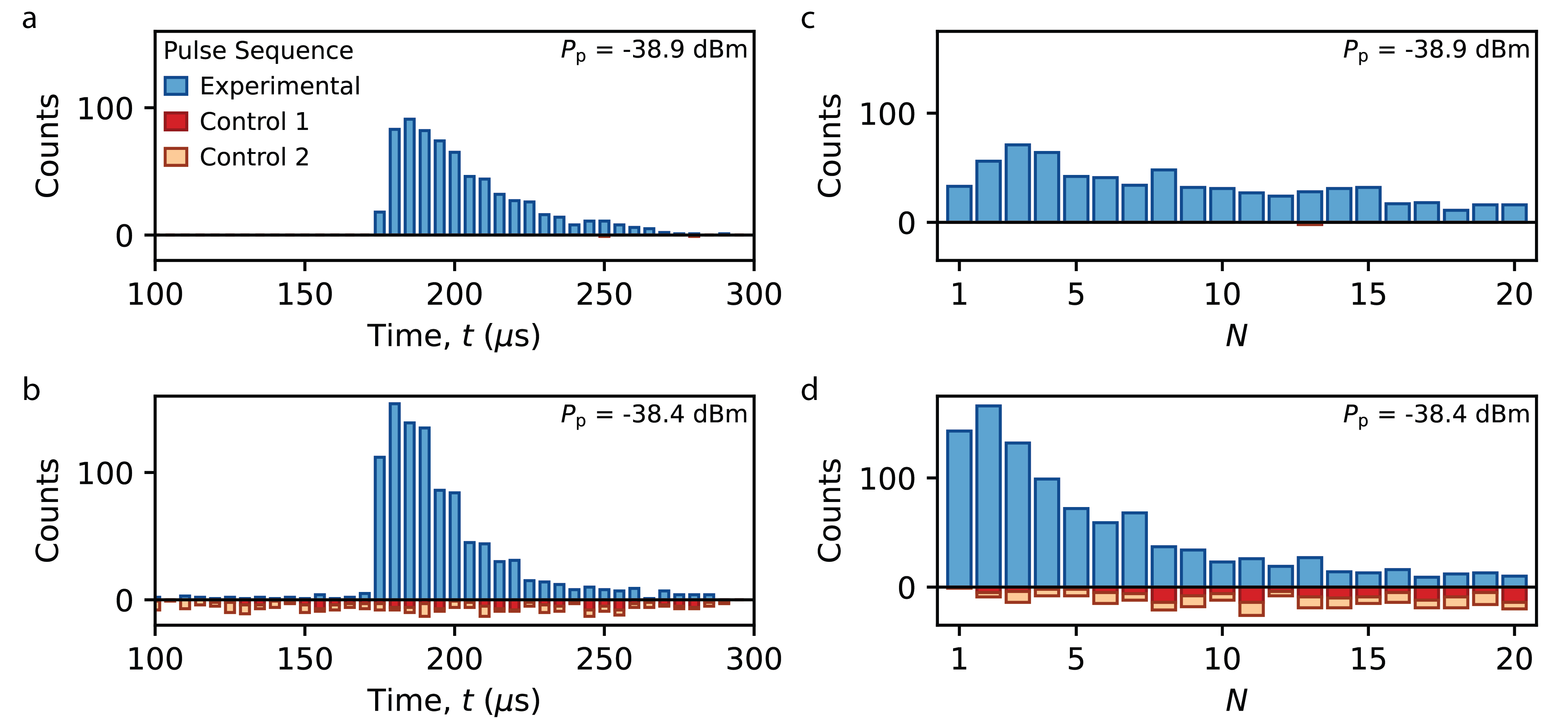}
	\caption{Histograms of detector ``clicks'' for a CPMG measurement. (a,b) The distribution of ``clicks'' with time $t$ for two $\Pp$. The counts associated with the two control experiments are summed and plotted as negative values, so that they can be easily compared with the counts from the experimental sequence. (c,d) The distribution of clicks with refocusing pulse repetition $N$. The data is from the experiment shown in Fig.~4c of the main text.}
	\label{fig_SI:cpmg_hist}
\end{figure}

In Figs.~\ref{fig_SI:cpmg_hist}a,b we plot histograms showing the time within the CPMG pulse sequence at which the detector ``clicks.'' The data is from the same experiment shown in Fig.~4c of the main text where $N=20$ refocusing pulses were used, and corresponds to the minimum and maximum values of $\Pp$ used in that experiment. For each shot of the pulse sequence, we consider only the first time the detector ``clicks,'' because the amplitude of the self-oscillating state is sufficiently large to drive the spin system and scramble the echoes. For both $\Pp$, the ``clicks'' of the experimental pulse sequence (blue bars) are triggered within a short time window near $t=170~\mu$s, corresponding to the time at which the spin echo refocuses (see the inset of Fig.~4b of the main text). In contrast, the dark counts for the two control experiments (red and orange bars, plotted as negative values) show no trend with $t$. These histograms give further confidence that the ``clicks'' triggered in the experimental pulse sequence are indeed caused by spin echoes and are not an artifact of the measurement sequence.

In Figs.~\ref{fig_SI:cpmg_hist}c,d we compare the distribution of the ``clicks'' with the repetition $N$, i.e. on which of the refocusing $y_\pi$ pulses the detection is made. The data clearly reveals that for larger $\Pp$, the ``clicks'' occur more frequently at smaller $N$. This is intuitive, as it suggests that as $\Pp$ approaches $\Pth$, the sensor is more likely to ``click'' on any given repetition, in accordance with the model described in Section~\ref{sec:EvsN}.

\section{Scaling of Spin Detection Efficiency with $N$}
\label{sec:EvsN}

\begin{figure}
	\centering
	\includegraphics[width=89mm]{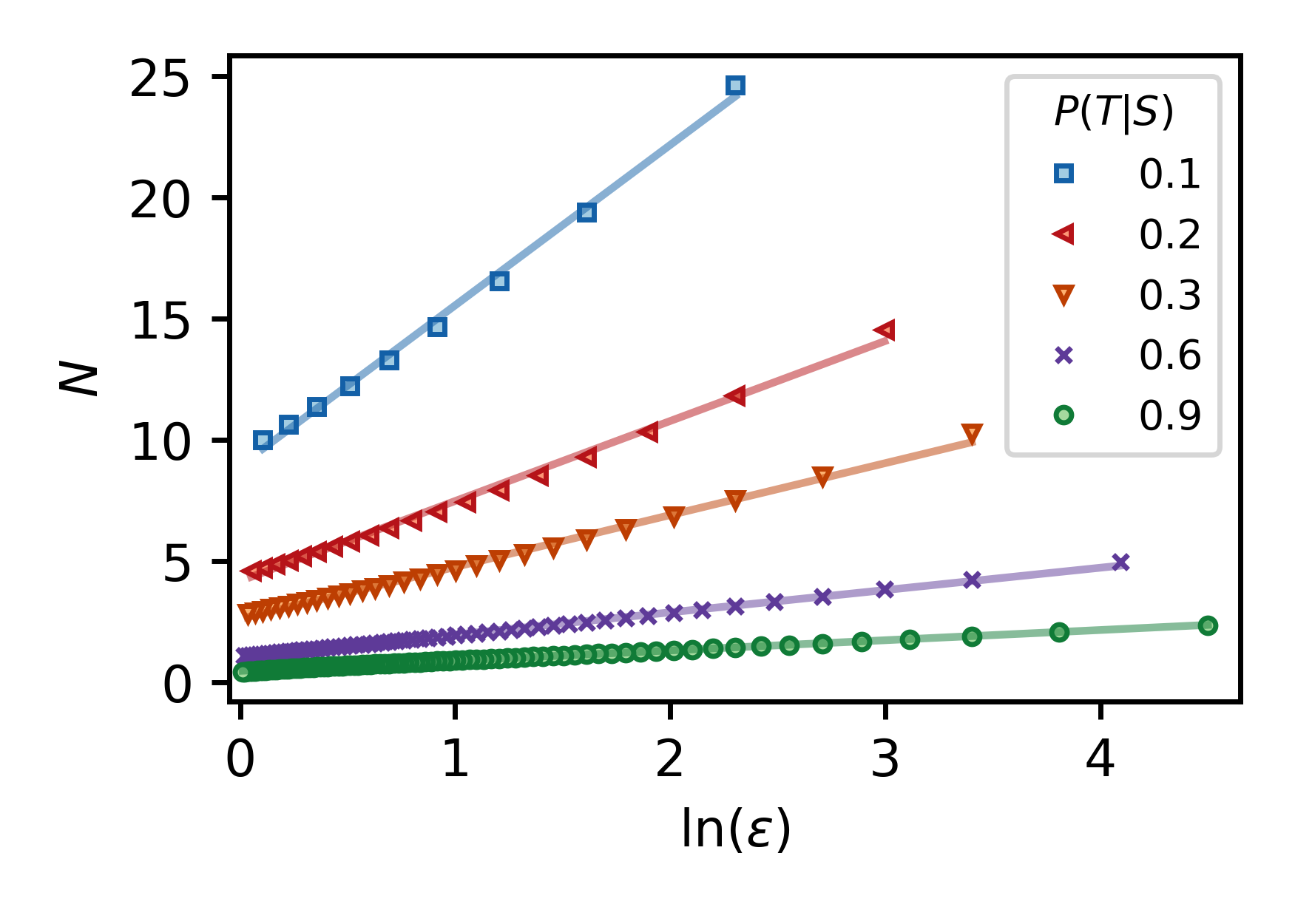}
	\caption{The number of refocusing pulses $N$ that maximizes the detection efficiency $E$ in a CPMG experiment. The points are calculated using Eq.~\ref{eqn_cpmg_scaling_N}. The solid lines are linear fits of $N$ as a function of $\ln(\epsilon)$.}
	\label{fig_SI:cpmg_scaling_N}
\end{figure}

In Figs.~4c,d of the main text we demonstrate that the detection efficiency $E$ of a spin echo signal obtained with a CPMG pulse sequence employing $N$ refocusing pulses can be modeled by the equation

\begin{equation}
    E(N) = [1-\Pdark]^N - [1-\Pbright]^N
    \label{main:equation_efficiency_scaling}
.\end{equation}

\noindent Here $\Pbright$ and $\Pdark$ are the probabilities associated with measuring a ``click'' with the experimental and control pulse sequences in a Hahn ($N=1$) echo experiment, respectively. From this equation we can calculate the $N$ at which $E$ is maximized by setting $\partial E(N)/\partial N = 0$ and solving for $N$. This yields

\begin{gather}
    \frac{\partial E(N)}{\partial N} = [1-\Pdark]^N \ln[1-\Pdark] - [1-\Pbright]^N \ln[1-\Pbright] = 0
    ,\\
    N = - \frac{\ln\left(\frac{\ln\left[1-\Pdark\right]}{\ln[\Pbright]}\right)}{\ln[1-\Pdark] - \ln[1-\Pbright]}
    \label{eqn_cpmg_scaling_N}
.\end{gather}

In Fig.~\ref{fig_SI:cpmg_scaling_N} we plot Eq.~\ref{eqn_cpmg_scaling_N} as a function of the ratio $\epsilon = \Pbright/\Pdark$ for several fixed values of $\Pbright$. The solid lines are linear fits of $N$ as function of $\ln(\epsilon)$, and capture the points well. This shows that the number of refocusing pulses $N$ in a CPMG experiment required to maximize the detection efficiency $E$ scales as $\mathcal{O}[\ln(\epsilon)]$.

\clearpage
% \printbibliography
\bibliographystyle{unsrtnat}
\bibliography{sci_adv/Self_Oscillation_Paper}